# Beyond Benchmarks: Evaluating Large Language Models and Agentic Patterns for Static Code Analysis Warning Classification

Lakmal Deshapriya, Sherlock A. Licorish, Brendon J. Woodford, Grant Dick

*School of Computing*
*University of Otago,*
*Dunedin, New Zealand*
lakmal.deshapriya@postgrad.otago.ac.nz, sherlock.licorish@otago.ac.nz,
brendon.woodford@otago.ac.nz, grant.dick@otago.ac.nz

***Abstract:***

***Context:*** *Static Code Analysis (SCA) tools support software verification and validation, reducing the time and cost of code reviews. However, the best SCA tools in the practitioner community generate a high volume of false positives, which developers tend to ignore, leading to genuine warnings being overlooked. Accordingly, studies increasingly adopt Large Language Models (LLMs) to identify and reduce false positives. Existing studies have produced promising results on public datasets, but underperformed on unseen datasets. Moreover, researchers do not leverage the full potential of Agentic patterns in LLMs or incorporate established code quality standards for warning classification.* ***Objectives****: This study utilises four manually annotated datasets to evaluate the performance of three contemporary LLMs for SCA warning classification in real-world applications, and incorporates Agentic patterns and established coding standards to enhance models' performance, addressing gaps identified in existing studies.* ***Method****: Three LLMs were selected using the LiveCodeBench benchmark, and the Structured Chain-of-Thought (SCoT) prompting technique was used to evaluate these LLMs. Furthermore, we incorporated Agentic patterns, such as Debating and Self-Consistency, to potentially improve performance in SCA warning classification.* ***Results****: Our results show that the performance of different LLMs varies with the characteristics of the dataset, ranging from 0.52 to 0.96 F1-Score for the best models. Additionally, the use of Agentic patterns does not improve LLMs' performance in most scenarios.* ***Conclusions****: SCA warning classification using LLMs works best only with certain warning types. We recommend evaluating LLMs in real-world applications rather than using publicly available datasets/benchmarks to understand their full utility.*

# 1. Introduction

Static Code Analysis (SCA) tools play a crucial role in software verification and validation. However, according to existing studies, SCA tools exhibit high false positive[1] warning rates, wasting developers' time and effort in adjudication [1, 2]. Unfortunately, an excessive number of false-positive warnings may lead developers to ignore all warnings [1, 3]. Thus, considerable effort is invested to reduce false positives by classifying or ranking SCA warnings [1, 4]. Nonetheless, these efforts do not fully leverage recent AI advancements, particularly Large Language Models (LLMs), for classifying SCA warnings [4-6]. LLMs are trained on extensive collections of natural language and software source code, enabling them to perform tasks such as code understanding and summarisation, debugging, code generation, test case generation, and documentation [7-9]. By utilising those capabilities, several previous studies utilised LLMs for SCA warning classification [2, 4, 6, 10]. However, multiple gaps in these studies prevent these models from being practically applied in software engineering.

For instance, as summarised in Appendix A, existing studies observed good classification performance on publicly available datasets. However, many frontier LLMs may already have been trained on these publicly available datasets, so these LLMs may classify SCA warnings by memorising what they have already seen rather than generalising knowledge to classify them [6, 11, 12]. Despite higher performance in existing studies, implementing these approaches to classify real-world SCA warnings[2] may yield unsatisfactory results. Conversely, different companies and software project teams may classify SCA warnings differently as true or false positives according to their individual preferences [13, 14]. Some teams may use established standards, while others might use a different classification based on their personal experience. Therefore, software developers must choose the appropriate standards based on their preferences over the suggestions generated by LLMs. Furthermore, advanced prompting and decoding techniques such as "Debating" and "Self-Consistency" require further evaluation across different real-world datasets [15]. Therefore, it is worth considering advanced techniques to improve SCA warning classification performance.

To address these gaps, two research questions were defined. The first research question was defined to address the first two gaps, using our own manually annotated SCA warning datasets, incorporating defined standards to classify warnings [16, 17]. The second research question was defined to address the third gap outlined in this section, aiming to improve the performance of SCA warning classification by integrating Agentic Patterns. The defined RQs are as follows,

1. What is the performance of pre-trained models in classifying SCA warnings using established coding standards?

[1] In this study, we used the 'false positives' for two contexts. First, the incorrect warnings generated by SCA tools were called false positives, and these warnings were labelled as "1" in the dataset. The second context is that, when analysing the performance of our approach, LLMs classified some warnings as "1", but their actual label is "0". Those incorrectly classified data points were also called false positives when experimenting with LLMs.

[2] SCA warnings that were not seen previously by LLMs

2. How do the Agentic Patterns (Debating and Self-Consistency) elevate the performance of the pre-trained best models in classifying warnings generated by SCA tools?

Our study makes several contributions to academia and the practitioner community as follows.

- We evaluated the three best LLMs based on the LiveCodeBench benchmark[3] across our four datasets, which have not been seen by LLMs before. Thus, we are the first to systematically demonstrate LLMs' performance on unseen datasets.
- We demonstrate how to incorporate company/individual coding standards when classifying SCA warnings by LLMs. This facilitates the practitioner community in adopting their defined standards for SCA warning classification rather than using standards that LLMs have seen before.
- We demonstrated that, despite using advanced prompting and decoding techniques, such as Self-Consistency and Debating, the performance of SCA warning classification does not improve when the dataset is unfamiliar to the LLM.

The rest of the study is organised as follows. Section 2 discusses the background of the study, and Section 3 reviews the existing studies related to the use of LLMs to classify SCA warnings. Section 4 outlines the datasets used in our study. Furthermore, Section 5 documents the methodology followed, and Section 6 summarises the results obtained. We then discuss our results and their implications in Section 7. In Section 8, we outline potential threats to the validity of our outcomes. Finally, Section 9 concludes the work and highlights potential areas for future research. Moreover, we provide a replication package containing the raw experimental results and the complete source code used to analyse the LLMs, enabling full reproducibility of our findings [18].

# 2. Background

## 2.1. Performance metrics used in this study

Several performance metrics, such as MCC, F1-Score, Balanced Accuracy and the time taken to analyse an SCA warning were utilised to evaluate the performance of LLMs in this study. **The Matthews Correlation Coefficient (MCC)** is a comprehensive performance metric for evaluating binary classifiers, especially when the dataset is imbalanced [19]. The MCC uses all four categories of the confusion matrix: true positives, true negatives, false positives, and false negatives to measure the performance, making it valuable when working with imbalanced datasets. On the other hand, metrics such as accuracy can be deceptive and yield misleading insights into the performance of SCA warning classifiers. A MCC score of +1 indicates perfect predictions, while zero signifies performance no better than random guessing. Moreover, the MCC score of -1 represents complete disagreement between predictions and actual outcomes. In summary, MCC reliably indicates model effectiveness, serving as an essential tool for researchers and practitioners in machine learning and statistical analysis [20].

[3] LiveCodeBench is a leaderboard of LLMs for code, which can be found in https://livecodebench.github.io/leaderboard.html

The F1-Score is a statistical performance metric used to evaluate a binary classification model. It is the harmonic mean of Precision and Recall, balancing the two metrics. Here, Precision measures the accuracy of positive predictions, while Recall (or Sensitivity) measures the model's ability to identify all relevant instances [21, 22]. The F1-Score varies from 0 to 1, with 1 representing perfect Precision and Recall, and 0 indicating the lowest performance. It is a valuable metric in situations with class-imbalance, as it accounts for both false positives and false negatives [22].

**Balanced Accuracy** is a better performance metric than accuracy, as accuracy can be misleading when one class dominates the dataset. However, Balanced Accuracy solves this issue by averaging Sensitivity and Specificity. Sensitivity measures how effectively the model detects the positive class, and specificity assesses how well the model identifies the negative class. Consequently, Balanced Accuracy measures the performance of the binary classification model in terms of correctly classifying both true positives and false positives, making it especially helpful for assessing models on imbalanced datasets, where one class may dominate the other [20]. By giving equal attention to both classes, Balanced Accuracy ensures that the model performs well across all categories, not just the majority class [20].

The **time taken to analyse an SCA warning** is the average time per analysis. This metric measures the latency of generating a prediction for an SCA warning. It helps select suitable LLMs and decoding methods to develop SCA warning classifiers, particularly when dealing with large datasets or real-time systems. This measure helps balance performance in terms of metrics such as F1-Score, Balanced Accuracy, and MCC with the limitations of available computational resources. Consequently, classifiers that take less time to analyse an SCA warning would be preferred.

### 2.2. Purposes of building a classifier for SCA warnings

A comprehensive literature study was conducted in Section 3. However, none of the existing studies considered the potential purposes of building a classifier for SCA warnings. Instead, existing studies have only tried to evaluate their approaches using the percentage of false positive warning reduction [2, 4, 5, 10, 23-25]. Therefore, we identified four primary goals for this study to classify warnings generated by SCA tools. Additionally, we selected suitable performance metrics to evaluate these goals (Refer to Section 2.1).

The **first goal** is to maximise the overall performance of SCA warning classification. We can assess the overall performance of LLMs in classifying SCA warnings using metrics such as accuracy, which offers a solid understanding of how effectively all warnings are classified. However, the datasets used in this study are biased towards actual or false positive warnings. Therefore, relying solely on accuracy provides less reliable insights into overall performance. As a result, we employed the MCC performance metric to evaluate the effectiveness of the SCA warning classification task, since this study involved binary classifications using LLMs.

The **second goal** of using LLMs to classify SCA warnings is to detect and reduce false positive warnings as much as possible. This is the most frequently discussed goal in existing studies [2, 4, 6, 10]; however, other goals also exist that can be achieved in similar studies. This helps to lessen developers' frustration with identifying false positive warnings. However, LLMs that identify as many false positive warnings as possible should not cause actual warnings to be misclassified as false positive warnings. In our dataset, false positive warnings are labelled as “1” and actual warnings as “0”. Therefore, we need to find an approach that delivers excellent performance on a metric that maximises true positives (False positive warnings in our dataset) and minimises false positives (Actual warnings in our dataset) and false negatives (False positive warnings misclassified as actual warnings). In this context, reducing false positives is mandatory, and ideally, it would be preferable to have zero false positives in an ideal situation. Because a model would classify actual warnings as non-actionable, developers will ignore these warnings without further inspection. Conversely, reducing false negatives is also important, as it would prevent false positive warnings from being misclassified as actual warnings, thereby saving the efforts of developers from unnecessary investigation. However, minimising false negatives is less critical than reducing false positives. Considering these factors, the F1-Score is the most suitable performance metric for the second goal. The F1-Score is the harmonic mean of Precision and Recall [11]. When the F1-Score reaches 1.0, all false positive warnings are correctly identified as false positives, and all warnings classified as false positive warnings are actually false positives. In this ideal scenario (F1-Score = 1.0), developers can disregard all the identified false-positive warnings and focus on the remaining ones. Therefore, if the aim is to optimise the classifier for this goal, the objective is to find the best approach or LLMs with the highest F1-Score. Consequently, developers can eliminate all warnings classified as positives, though some actionable warnings might also be removed, especially if the F1-Score is not perfect.

Practically, what developers need is to identify actionable warnings. For this, we can identify false positive warnings similar to the second goal and remove them from the entire set of warnings to obtain actionable warnings. However, the second goal was created to maximise the false positive warnings, which include some actionable warnings. Therefore, we defined the **third goal** to identify the maximum possible actual warnings with a minimum number of false positive warnings, where the loss of actual warnings is minimised. However, false positive warnings may still occur. Considering the confusion matrix, the aim is to maximise true negatives while minimising false positives and false negatives. In this context, Balanced Accuracy (BA) can be employed as a performance metric, taking into account both the True Positive Rate (TPR) and the True Negative rate (TNR) [20]. Using TPR and TNR makes BA especially valuable in situations where the classes are imbalanced. In the ideal scenario where the BA value reaches 1.0, the classifier accurately identifies all actual warnings. Moreover, we do not use Accuracy as a performance metric for this purpose, as our datasets are imbalanced towards either false positives or true positives. In such a scenario, BA is better for class-imbalanced datasets [20]. That said, developers can reduce the time spent on investigating false positive warnings.

In addition to the other goals discussed in this section, the **time required to classify a single SCA warning** is a practical consideration when implementing the classification model in a production environment. Therefore, we tracked the time taken to classify SCA warnings for each dataset and LLMs separately. All the goals discussed in this section are summarised in Table 1.

*Table 1 Goals for model building*

| Goal ID | Purpose of model building | Metric used |
|---|---|---|
| G1 | Maximise the overall model performance | MCC |
| G2 | Correctly detect false positive warnings as much as possible to minimise false warnings | F1-Score |
| G3 | Correctly identify the maximum possible actual warnings with minimum false positive warnings | Balanced Accuracy |
| G4 | Reduce the time taken to process an SCA warning | Time in seconds |

### 2.3. LLMs optimisation with fine-tuning, hyperparameter optimisation and prompt engineering

LLMs can be optimised using various approaches, including fine-tuning, hyperparameter optimisation, and prompt engineering [26-28]. These different optimisation approaches focus on different aspects of the model. Model fine-tuning involves specialising LLMs for specific tasks, such as source code generation, summarisation, and solving maths questions [29]. However, fine-tuning demands substantial resources and large specialised datasets [30]. Therefore, we did not incorporate fine-tuning for our study. The second method for optimising LLMs is hyperparameter optimisation, which can improve LLM performance without altering the models. Several hyperparameters are relevant when prompting LLMs, including Temperature, Top Probability (top_p), Frequency Penalty, and Presence Penalty [31]. The temperature parameter primarily controls the randomness of the output [31]. Top Probability (top_p) restricts the sampling pool of tokens, limiting the allowed tokens for the next predicted word [31]. Since this study utilises LLMs for binary classification, setting the temperature to a lower level of randomness (temp = 0.2) and setting top_p between 0.75 and 1 (top_p = 0.95 in this study) is advantageous for making the response more factual rather than being creative [31]. Moreover, the Frequency Penalty discourages repeated tokens and encourages more novel outputs [31]. Conversely, the presence penalty reduces the model from generating tokens that have already been generated, promoting diversity in the generated output [31]. Because the model output is expected to include the predicted label (true positive or false positive), a reason for the decision, and any missing details, we set the max_tokens parameter to 16384, because it matches a common 16k context window of LLMs, giving the model room to generate long, uninterrupted outputs without being cut off[4] [32]. It should be noted that only the temperature, top_p, and max_tokens hyperparameters significantly influence models' performance, while other hyperparameters will have a minimal effect on SCA warning classification [31].

Similar to hyperparameter optimisation, prompt engineering is also used to enhance the performance of LLMs without altering internal model parameters [33]. It can be described as the process of designing or shaping natural language instructions to achieve desired behaviours and outputs from pre-trained language models [34]. Instead of updating model weights through fine-tuning, prompt engineering allows seamless integration of pre-trained models into downstream tasks by providing task-specific instructions, called prompts, that guide the model's response generation [33]. Several techniques have been developed for prompt engineering, including Zero-shot and Few-shot learning, as well as Chain-of-Thought (CoT).

[4] The concept of "Context window" is explained here: https://cobusgreyling.medium.com/what-does-the-openai-16k-context-window-mean-a0d2e10f7bfa

According to Google's prompt engineering guidelines[5], the Zero-shot prompting technique involves instructing an LLM with a direct request without providing any examples [33]. With this approach, we must depend solely on the LLM's internal knowledge, unless the relevant information is provided with the prompt. However, we can utilise One-shot, Few-shot, or Multi-shot prompting methods to present one or more examples of desired input-output pairs to the LLM, enhancing its understanding of the specific task and resulting in more accurate responses [33]. Moreover, the CoT prompting technique encourages models to generate intermediate reasoning steps before arriving at a final answer, significantly improving performance on complex reasoning tasks [34]. With CoT, the original problem is decomposed into smaller sub-problems, enabling each to be solved effectively. The results of each sub-problem are then combined to produce the final solution. There are automatic versions of CoT as well, by adding a string similar to "Let's think step-by-step" to the prompt, which leads the LLM along a CoT generation path without user intervention [33].

Furthermore, there are other prompting techniques, including Self-Consistency and Generated Knowledge [34]. The Self-Consistency prompt engineering method generates multiple reasoning paths and selects the most consistent answer, enhancing the reliability of model outputs [34]. Moreover, the Generated Knowledge approach involves prompting the model to first generate relevant background knowledge before answering the main query, thereby improving factual accuracy [34]. Other prompting techniques are also noted in other works [33-35]. These do not train the underlying parameters of the LLM, but instead modify the input data and the way calculations are performed in the LLM, resulting in significantly different outcomes. Therefore, the selection of the prompting technique is vital [33].

LLMs are algorithms that predict the next token based on the decoding methods used. There are multiple decoding methods available in LLMs to generate outputs, such as Greedy Search, Beam Search, Temperature Sampling and Top-p Sampling (Nucleus Sampling) [36]. Greedy Search selects the next token with the highest probability. However, this can lead to a local optimal solution, missing the global optimal solution. To address this issue, the Beam Search approach is introduced. Beam Search improves the Greedy Search by keeping track of the $k$ most probable sequences (Beams) at each step. It explores a broader range of possibilities than greedy search. However, it can still get stuck in local optima and lacks diversity in open-ended generation tasks. In contrast, the Temperature Sampling technique adjusts the sharpness of the entire probability distribution over tokens using a scalar parameter called temperature (T). Higher temperatures (e.g., $T > 1$) flatten the distribution, increasing the likelihood of selecting lower-probability tokens, leading to more diverse and creative outputs. Lower temperatures (e.g., $T < 1$) make the distribution sharper, favouring high-probability tokens and resulting in more deterministic and focused text. Moreover, Top-p Sampling (Nucleus Sampling) is an adaptive decoding method that selects the smallest set of tokens whose cumulative probability exceeds a predefined threshold p (e.g., 0.9). This allows the number of considered tokens to vary dynamically based on the probability distribution's shape, providing flexibility and generally producing more natural-sounding text than fixed top-k sampling.

[5] Google explains multiple prompting techniques. More information can be found here: https://cloud.google.com/discover/what-is-prompt-engineering

This study aims to classify SCA warnings into actionable and non-actionable categories. For that, the LLM should understand the source code (or the relevant slice of it) and the SCA warning and be able to identify the relevant coding standards to produce the correct response. Since this is a classification task and the output should be factual, the recommended hyperparameters were used to maximise the LLM's performance, with Temperature Sampling and Nucleus Sampling decoding methods. Additionally, for this task, information such as coding standards and rules defined by SCA tools is involved. Fine-tuning LLMs with this information is a good option; however, for that, there should be a significantly large dataset [30]. Therefore, we use existing fine-tuned models with a large set of open-source coding repositories, which improved source code understandability. Consequently, in this study, we aim to improve the performance of SCA warning classification through optimising prompt engineering.

### 2.4. Providing external knowledge using RAG and CAG

Retrieval Augmented Generation (RAG) is utilised to supply external information to the LLM model. In RAG, data relevant to a specific task is searched from a vector database, pre-processed, and presented with the LLM prompt. RAG assists LLMs in accessing current and factual data, particularly for the task. LLMs trained with data available before the cut-off date and data generated after that date are excluded from the LLM [37]. Consequently, LLMs can produce inaccurate results, and their outputs are outdated. To address this, RAG is employed. However, during the data search stage, RAG might retrieve irrelevant data, increase the token size of the prompt and reduce the accuracy of the generated results [38]. To mitigate this issue, we used Context Augmented Generation (CAG) [39], where only the related data is retrieved based on the type of the SCA warning assessed. Consequently, only the relevant data will be fed to the LLM along with a slice of the applicable source code and the SCA warning message.

### 2.5. AST graphs and Data dependency graphs

Abstract Syntax Tree (AST) is used to represent the structure of software source code in a tree-like form, where each node in AST represents a particular construct, such as classes, methods, and variables [40]. AST represents only the structural and content-related details of the source code, while discarding any additional details, such as comments. Information that gets transferred to the AST includes:

- The variable types and their declaration locations
- The order and definition of executable statements
- The left and right components of binary operations
- Identifiers and their assigned values

AST graphs are used in this study to extract program slices from source code where the relevant lines are spread throughout, and the relationship between these lines depends on their structure. In contrast, a data dependency graph is a directed graph showing the dependencies between data items in source code [41]. For example, within source code, a variable definition with an initial value creates data dependencies through the modification of that variable and its subsequent use. In this study, data dependency graphs are utilised to extract code slices from source code, where relevant lines are dispersed throughout the code and their relationships depend on the data or identifiers/variables.

# 3. Related work

In prior studies, several efforts have been made to classify SCA warnings using LLMs, and this section reviews these approaches. Kharkar et al. [23] used approaches, such as a feature-based model utilising handcrafted features, DeepInferEnhance (a fine-tuned CodeBERTa model on Java and C# functions), and GPT-C (a variant of the GPT-2 code completion model in a Zero-shot setting) to classify SCA warnings. Using a dataset consisting of 539 null dereference warnings from open-source and proprietary projects, they demonstrate that DeepInferEnhance improves Precision by 15.13% and GPT-C achieves the best Precision gain (+17.47%) without additional training. Additionally, the approach generalises to resource leak warnings, where GPT-C improved Precision by 5.56%. In this study, Kharkar et al. [23] demonstrated the potential for reducing the false positive rate in the SCA warning dataset. However, the primary drawback of this study is that the Recall value decreases from 100% to 83.7% in GPT-C when compared to the baseline.

Ge et al. [24] systematically examined the feasibility of using Pre-Trained Models (PTMS), which are trained on billions of code tokens for SCA warning classification. They assessed five representative PTMs, including CodeBERT, GraphCodeBERT, CodeT5, UniXcoder, and CodeGPT. These PTMs encompass encoder-only, decoder-only, and encoder-decoder architectures. The evaluation was conducted on a dataset comprising over 10,000 SpotBugs warnings from 10 large-scale Java projects. All studied PTMs significantly surpass the State-Of-The-Art (SOTA) deep learning-based Actionable Warning Identification (AWI) approaches, such as CNN and LSTM. PTMs enhanced performance in terms of AUC by a margin of 9.85% to 21.12% over the SOTA baseline models. The CodeBERT model (an encoder-only architecture) achieved the best performance with an AUC of 70.77%. The study indicates that encoder-only PTMs are most suitable for this code classification task compared to encoder-decoder or decoder-only models. The authors of the study provided several practical guidelines to improve actionable warning identification systems, including reducing the context window by minimising the size of source code using techniques such as program slicing, and enhancing AWI by training domain-specific PTMs and fine-tuning these PTMs. Additionally, Ge et al. [24] suggest using a training dataset with diverse patterns SCA warnings to improve AWI performance. This will help to train the AWIs with diverse patterns.

Flynn and Klieber [5] presented an approach for leveraging LLMs, such as GPT-4, to automate the adjudication of Static Analysis (SA) alerts. The authors developed a model where an LLM-based tool ingests source code and SA alerts, then queries the LLM to output adjudicated true positives with a trace, false positives with a proof sketch, or “uncertain”. To mitigate the limitation of the context window, they employed several approaches. One approach is using traditional SCA tools to produce the required information, and another is using the LLM itself to generate function summaries. Additionally, they instructed the LLM in the prompt to ask for the needed information to prevent the limitation of the context window. Furthermore, they used the LLM to create preconditions to avoid a bad state in a function, then had the LLM check whether callers satisfy the precondition. Although the paper did not provide statistical results, it demonstrated several examples of GPT-4 successfully performing complex tasks related to alert adjudication, with historical LLM chat records.

Apostolidis et al. [25] proposed a novel approach to reduce non-actionable warnings generated by CppCheck. They analysed the same software code using both CppCheck and fine-tuned CodeBERT LLM to predict vulnerabilities. If CodeBERT predicted a function as vulnerable, CppCheck warnings within that function were retained as actionable; otherwise, they were classified as non-actionable warnings. They could improve the Precision (from 5.88% to 96.20% in the Big-Vul dataset and from 8.71% to 47.67% in the ReVeal dataset), while Recall decreased from 46.90% to 40.66% on Big-Vul and from 6.88% to 5.93% on ReVeal. Thus, higher Precision was achieved while some genuine detections were lost. However, in this study, SCA warning details were not incorporated into the LLM for its predictions, and CodeBERT classified functions rather than individual warnings. A vulnerable function may contain both actionable and non-actionable warnings, while an incorrect prediction that a function is safe may remove actionable warnings. Additionally, the datasets were collected from public commits linked to known vulnerabilities. Such labels do not confirm the correctness of every SCA warning in those functions. Therefore, Apostolidis et al. [25] provide evidence for selecting functions for review, but more limited evidence for classifying individual warnings.

The previously mentioned attempts do not completely harness the full potentials of LLMs in classifying SCA warnings. These attempts utilised Small Language Models or pre-trained transformers to classify SCA warnings. Additionally, some of them used SCA tools and LLM as an ensemble to analyse code independently, without cooperating with each other [25]. Wen et al. [2] introduced LLM4SA, the first LLM-based SCA warning classifier that analyses both source code and SCA warning messages with an LLM. They used LLMs such as, ChatGPT and Llama-2 and SCA tools such as CppCheck, FB Infer, and Clang Static Analyser. To overcome the token window limitation when analysing large codebases, they employed static program-dependence traversal to extract the relevant slice of the source code, providing sufficient context related to bugs [42]. Additionally, they improved prompting through Few-shot prompting and CoT. Their evaluation of the Juliet benchmark programs, three embedded operating systems and 11 real-world C/C++ projects showed that LLM4SA achieved high Precision (81.13%) and Recall (94.64%) for a total of 9,547 bug warnings. Additionally, they experimented with their approach on real-world applications, and the results were lower than the Juliet benchmark. However, the embedded-system evaluation contained mainly false warnings. Of the 749 warnings, only 13 warnings were actionable. The reported 100% Recall therefore provides limited evidence that actionable warnings would be preserved in other projects. In contrast, approximately 84% of the Juliet warnings were actionable, making high Precision easier to obtain even when some false warnings were retained. These different class distributions limit comparisons between the reported scores. They discussed possible data leakage but did not establish whether the evaluated code had been used in LLM training. If the LLM had seen the same code before, its performance could partly reflect memorisation rather than classification of unfamiliar warnings. Additionally, the study covered six bug categories using category-specific examples, so the results do not establish similar performance for other warning types. LLM4SA incorporated the bug report generated by the SCA tools to provide contextual information. However, they did not use any software source code quality standards or the relevant description of the warning given by the SCA tool in the official documentation. Without this contextual information, the LLM will classify SCA warnings based on its internal knowledge, which might be outdated or incorrect. Moreover, Svärd and Teräs [10] proposed a Large Language Model Assisted Static Analysis (LLMAS) prototype that integrates an LLM (Google Gemini) with an SCA tool (SpotBugs) to identify and reduce false positive warnings generated by this SCA tool. They used 10,056 test cases from the Juliet Java 1.3 dataset, covering 14 of the dataset's 112 Common Weakness Enumerations (CWEs). Svärd and Teräs [10] benchmarked LLMAS against the standalone SpotBugs, and the results showed that LLMAS improved Precision by 13.50%, from 0.3236, while maintaining the same Recall (0.4914). The selected CWEs also had unequal numbers of test cases, so common warning types may have a greater influence on the overall results. Juliet is also synthetic, and names such as "good" and "bad" may reveal the intended labels. These clues and possible training-data overlap may make the results more favourable than those obtained on unfamiliar real-world code. However, using LLMAS added significant latency, with execution times rising from 7.2 minutes (for SpotBugs) to 4.86 hours (for both SpotBugs and LLMAS), which may limit its use in Continuous Integration/Continuous Development (CI/CD) pipelines when rapid feedback is required.

In contrast, Chen et al. [4] identified several drawbacks in existing studies, including (1) imprecision, where LLM4SA utilises overly broad code snippets such as entire function bodies, cluttering inputs with irrelevant code and overwhelming the LLM, leading to inaccurate assessments, longer processing times, and higher token costs; and (2) incompleteness, where LLM4SA misses essential code context, such as definitions of global variables or external functions, which are vital for accurate data and control flow analysis. To address these issues, Chen et al. [4] proposed LLM4FPM, a lightweight and efficient framework for SCA warning classification. The primary contribution of this study is to extract more precise and complete source code context for the LLM. They identified several challenges in extracting the most relevant code context, including (1) insufficient fidelity or accuracy of the existing program slicer; (2) missing some dependencies in Joern's Code Property Graph (CPG) [43]; and (3) inefficient inter-file dependency extractors. To tackle these, they introduced two components: eCPG-Slicer and Find All Referenced Files (FARF). The eCPG-Slicer performs line-level, precise code slicing by extending Joern's CPG and creating a structured report. Moreover, FARF identifies all necessary dependent files for a given warning, providing the eCPG-Slicer with complete inputs, thereby eliminating the need for costly whole-program analysis. The code slice extracted using eCPG-Slicer, and the SCA warning report are combined into a structured format and fed to a lightweight, open-source LLM (Qwen 2.5-32B). The LLM is guided using specialised techniques such as Expert Prompting [44], CoT reasoning, Few-shot examples, and a Self-Consistency mechanism (majority voting) to determine whether a warning is classified as a "real bug" or a "false alarm". The results from this study demonstrated that providing precise and complete context significantly improves SCA warning classification using LLMs. Additionally, LLM4FPM outperformed the current SOTA method, achieving an overall F1-Score of 98.92% on the Juliet dataset. However, approximately 85% of the evaluated warnings were actionable. Retaining most warnings can therefore give a high F1-Score for actionable warnings, so this score also needs to be considered alongside false positive removal. The evaluation covered seven CWEs using one LLM and specialised prompts, limiting the evidence for other warning types and models. The authors also acknowledged possible training-data overlap with Juliet, so performance on unseen code remains uncertain. Additionally, LLM4FPM was used to relabel the noisy D2A dataset, with 43 of 50 manually checked examples from its libtiff subset labelled correctly (86%). This small sample from one project does not establish the accuracy of the complete relabelled dataset. Furthermore, testing on eight real-world open-source C/C++ projects demonstrated that the framework successfully reduced false positives by over 85% in the manually checked sample. The system is also effective, with an average inspection time of 4.7 seconds per warning, showing its potential for integration into CI/CD pipelines under the tested hardware and execution settings. Moreover, they evaluated their approach with eight real-world code repositories as well. They analysed those code repositories with SCA tools and selected 50 samples randomly from the warnings generated by SCA tools. Two human experts confirmed that 43 of the samples were false positives. They could not annotate the other 7 warnings. Out of the 43 confirmed warnings 37 cases are classified as false positives by their approach. Since the manually checked real-world sample contained no confirmed actionable warnings, this evaluation does not show whether the approach would incorrectly remove actionable warnings. The seven unresolved cases also limit confidence in its performance on warnings that are difficult to annotate. Therefore, our study conducted an in-depth analysis on this using four manually annotated datasets considering actionable and non-actionable warnings.

On the other hand, Iranmanesh et al. [6] proposed ZeroFalse to reduce false positive warnings while preserving coverage. They experimented with 10 LLMs to classify SCA warnings, which belong to 10 CWEs, generated by CodeQL. They used CWE-specific guidance within a common prompt template to help LLMs understand the context better. Their best model achieved 0.912 in terms of F1-Score for the OWASP Java benchmark and 0.955 on the OpenVuln dataset. The OWASP evaluation contained 1,449 actionable warnings and 525 false positives, with no false positives for two CWEs. Results for these two categories therefore do not show how well the approach rejects false warnings. According to Iranmanesh et al. [6], CWE specialised prompts improved results for most models compared with prompts without information regarding CWEs. However, some models performed worse, so the benefit of this guidance depends on the model and dataset. Additionally, they found that reasoning models outperformed other LLMs. However, their OWASP evaluation covered only 10 CWEs, so the results do not establish similar performance for other warning types. Extending the approach requires preparing and validating guidance for those types. Furthermore, they used a common program slicing approach for different CWEs, but did not evaluate whether warning-specific slice would further improve the results.. Moreover, proprietary LLMs achieved the highest F1-Scores, although some open-weight models also performed well with this approach. Grok-4 and Gemini-2.5 pro were the best models for OWASP benchmark results, and GPT-5 and Grok-4 performed best in OpenVuln benchmark results. In this study, OpenVuln contained 58 warnings from seven open-source Java projects, including 23 actionable warnings and 35 false positives. The small sample limits conclusions about other projects. Its source code and known fixes were publicly available, so the study does not establish that they were unseen by the LLMs. Prior exposure could make the results less representative of unfamiliar code. Labels were assigned by comparison with known fixes, which may not identify other faults in the same code. Therefore, agreement with these labels does not confirm that every rejected warning is non-actionable. Their experiments shows that the OpenVuln dataset underperforms when compared to the OWASP dataset for multiple LLMs. Additionally, they mentioned the necessity of real-world dataset which is not trained for LLMs. In our study, we show that our four datasets provide a good benchmark for evaluating LLMs as they are not trained on LLMs and have different false positive percentages. On the other hand, for practical implementation, the confidentiality of the source code should be considered. To prevent sharing source code with external parties, approaches that work on open-source models, which can be deployed in company environments, should be used. Consequently, there is a necessity for better approaches which consistently produce exceptional results using open-source models.

Appendix A summarised the LLMs, datasets and the reported performance in existing studies and overall. Existing studies have used unnecessarily large source code or slices from the source code to feed LLMs, expecting better performance in SCA warning classification. However, with the increase in input token size, the throughput of the SCA warning classification system decreases, due to the transfer of more data and an increase in processing time in the LLM. Additionally, the performance of LLMs decreased with the rise in input token size, as shown by Levy [38], mainly due to hallucination in LLMs. LLMs also hallucinate due to the lack of necessary information, then attempt to guess the answer with limited data [45]. Moreover, sufficient effort was not invested in effectively using coding standards to judge SCA warnings accurately. Therefore, in this study, we aim to eliminate this gap by providing all the necessary information for LLM to classify SCA warnings.

In summary, existing studies demonstrated the potential for reducing the false positive warning rate in SCA warning datasets [4-6, 23, 25]. These studies showed that LLMs work better than existing SOTA [24]. However, in our study, we identified that LLMs produce lower performance for datasets which is not seen by the LLM previously. Moreover, only some existing studies utilised the full potential of LLMs, evaluating both source code and SCA warning messages to identify actionable warnings from the set of warnings generated by SCA tools. Studies that used both source code and SCA warning messages as input data achieved promising results in identifying actionable warnings [2, 4, 6, 10]. Additionally, we embedded established source code quality standards with LLM prompts in classifying SCA warnings, as software engineering companies and teams may need to use their own standards rather than using the standards trained by LLMs. Therefore, in our study, we used source code, warning description and established source code quality standards which are relevant for the warning type (SCA rules that warnings were generated).

Moreover, some of the existing studies used different approaches for program slicing to fit the source code and warning messages into the limited token window [2, 4]. However, Chen et al. [4] identified two issues with existing program slices used as inputs to LLMs, such as imprecision and incompleteness. By addressing these two issues, Chen et al. [4] could harness the full potential of LLMs for SCA warning classification. However, in our previous study, manually annotating four SCA warning datasets, we identified that the approach needed for program slicing differs based on the relevant SCA rule [46]. Therefore, we defined several program slicing approaches for individual SCA rules. Moreover, Chen et al. [4] used multiple prompting techniques such as Expert Prompting [44], CoT reasoning, Few-shot examples, and a Self-Consistency mechanism (majority voting) and achieved near-perfect results (99%) on the Juliet public dataset and correctly relabelled 43 of 50 manually checked examples from the libtiff subset of D2A (86%). On the other hand, we used the Structured Chain of Thought (SCoT) prompting technique for this study, as it facilitates the embedding of relevant source code quality standards to the prompt and guide the LLM to generate intended results following specified procedures and rules. Furthermore, we used Agentic patterns such as Self-Consistency and Debating to potentially improve the performance of warning classifiers, extending the body of evidence.

Furthermore, Iranmanesh et al. [6] experimented with 10 LLMs such as Grok-4, Gemini-2.5, GPT-5 and Grok-4  to classify SCA warnings and found that the use of warning information in prompts improved results for most models. Additionally, they found that reasoning models generated higher performance in SCA warning classification. . Extending this approach requires additional CWE-specific guidance, and the study did not evaluate whether customised program slicing for each warning type would improve the results. However, SCA tools introduce and deprecate warning types from time to time [46]. Therefore, . maintaining CWE-specific guidance may require additional effort as warning types change. In our study, we use a common prompt template with the relevant coding standards for each warning type. However, the program slicing approach should be defined based on the warning type. Furthermore, as a next step, we plan to automate this process through an expert model as mentioned in Section 9.

Overall, we found several gaps in existing studies which are addressed in the current study. Existing studies did not use sufficient contextual information for SCA warning classification. We used software source code quality standards and the description of the warning given by the SCA tool in the official documentation [2, 10]. We provided this information with each SCA warning classification request. Additionally, we experimented with our approach on real-world datasets that have not been seen by LLMs. Moreover, we tried to optimise the results using better prompting techniques, adding more context and using Agentic AI patterns. Furthermore, we evaluated an open-source model which may be deployed using an on-premise server, which eliminates privacy concerns.

# 4. Datasets

Four SCA warning datasets[6] [46] were created by analysing two source code datasets (the Stack Overflow dataset collected by Meldrum et al. [47] and the Apache Tomcat source code downloaded from GitHub on 8th December 2023) using two SCA tools (PMD and SonarQube).

[6] Creation of these four datasets are now published as a separate Journal paper with comprehensive details [46].

The reason for choosing the Stack Overflow and Apache Tomcat datasets is that these two source code datasets have been utilised in multiple existing studies and are recognised as reliable testbeds [48-53]. Moreover, the Stack Overflow dataset consists of code snippets, whereas Apache Tomcat included full-length source code, thereby creating diversity across the four SCA warning datasets. Additionally, PMD and SonarQube are the leading SCA tools within the software engineering practitioner community and have the lowest false positive rates [54-56]. Therefore, PMD and SonarQube were used to scan the two source code datasets, generating the four SCA warning datasets. As a result, the four datasets contain SCA warnings from both code snippets and full-length source codes. The dataset summary is shown in Table 2, where the number of specific rules of SCA tools that generated warnings (number of rules triggered in SCA tools for the source code) is indicated as "# of rules found". Here, a considerably large number of warnings (indicated as "# of alarms" in Table 2) were produced by each SCA tool through the analysis of each code dataset. Furthermore, due to the significant resources required for manually annotating the entire datasets, four sample datasets were created randomly [57], with a 95% confidence level and a 5% margin of error. Thereafter, the sample size of each SCA rule was determined proportionally to the number of warnings in the population dataset [57, 58]. When the proportion of the sample size for a given rule was a fraction, it was rounded up, increasing the total number of samples, which is slightly above the 95% level. The sample sizes for each dataset are listed as "Sample size" in Table 2.

Following the defined coding standards from our previous study [16], SCA warnings in the sample datasets were manually reviewed to identify whether they were true positives or false positives. The number of false positive warnings for each dataset is listed as "# of false positives in the sample" in Table 2. Furthermore, the number of rules that produced false positive warnings is indicated as "# of rules generated false positive" in Table 2, and the false positive percentage for each dataset is displayed as "False positive percentage" in Table 2. It is worth noting that most of the rules did not produce any false positive warnings, and the associated warnings were removed, resulting in a dataset size referred to as "# of samples for evaluation" in Table 2. The final false positive percentages of these evaluation datasets are listed as "False positive percentage in the final dataset" in Table 2. These datasets were then used to evaluate LLM models and generate results for each research question as described in this study.

After selecting samples randomly, the sample dataset 1 consists of 531 samples, covering all the rules available in the population dataset [59]. Sample Dataset 1 consists of 98 false positives, which were generated by 13 SCA rules out of a total of 188 rules. Similarly, sample Dataset 2 consists of 536 samples, including 72 false positives arising from 16 SCA rules out of 197 rules, which generated all the SCA warnings in dataset 2. Sample Dataset 3 contains 646 random samples from Dataset 3, including 30 false positives from 5 SCA rules out of a total of 310 rules, which generated all SCA warnings for Dataset 3. Finally, sample Dataset 4 consists of 631 samples, which are randomly selected from Dataset 4, including 44 false positives from 14 SCA rules out of 281 rules that generated all the SCA warnings for dataset 1. Moreover, the standards used to review and identify false positives manually in all the datasets are shown in Table 3, including the reason for selecting them for our study.

*Table 2 Summary of datasets*

| Dataset Name | SCA tool | Code Dataset | # of rules found | # of alarms | Sample size | # of false positives in the sample | # of rules generated false positives | False positive percentage | # of samples for evaluation | False positive percentage in the final dataset |
|---|---|---|---|---|---|---|---|---|---|---|
| Dataset 1 | PMD | SOD | 188 | 80788 | 531 | 98 | 13 | 18.45% | 139 | 70.50% |
| Dataset 2 | PMD | AT | 197 | 176293 | 536 | 72 | 16 | 13.43% | 117 | 61.54% |
| Dataset 3 | SonarQube | SOD | 310 | 83890 | 646 | 30 | 5 | 4.64% | 150 | 20.00% |
| Dataset 4 | SonarQube | AT | 281 | 128524 | 631 | 44 | 14 | 6.97% | 298 | 14.77% |

SOD - Stack Overflow dataset, AT - Apache Tomcat

*Table 3 Established standards*

| Id | Established Standards | Reason for selecting the standard |
|---|---|---|
| 01 | https://wiki.sei.cmu.edu | Wiki page maintained by Carnegie Mellon University and used as a standard for SonarQube |
| 02 | Book: Clean Code. | Used as a standard for SonarQube |
| 03 | Book: Effective Java by Joshua Bloch. | Used as a standard for SonarQube |
| 04 | Book: Mastering Regular Expressions by Jeffrey Friedl. | Used as a standard for SonarQube |
| 05 | Book: Refactoring: Improving the Design of Existing Code by Martin Fowler, Kent Beck, John Brant, William Opdyke, and Don Roberts. | Used as a standard for SonarQube |
| 06 | Google Java Coding Style Guide. | Used as a standard for SonarQube and A popular and well-accepted standard [60, 61] |
| 07 | https://cwe.mitre.org/data/definitions | Used as a standard for SonarQube and A popular and well-accepted standard [62, 63] |
| 08 | Java Doc. | Java official user guide |
| 09 | Book: Object-Oriented Metrics in Practice: Using Software Metrics to Characterize, Evaluate, and Improve the Design of Object-Oriented Systems by Michele Lanza and Radu Marinescu. | Used as a reference for PMD |
| 10 | Paper: Linguistic Antipatterns - What They Are and How Developers Perceive Them. | [64] |
| 11 | Principle of Least Astonishment. | [65] |
| 12 | SOLID principles. | Used as a standard for SonarQube |
| 13 | Transitive property of equality. | Used as a standard for SonarQube |
| 14 | https://owasp.org | A popular and well-accepted standard [60, 66] |

The sample datasets were manually annotated, with false positive warnings marked as "1" and true positive warnings marked as "0". These datasets were manually annotated in accordance with defined coding standards, with support from a software tool we developed to display all relevant information on one page. This mitigates mistakes that can occur while manually cross-checking source code, warning descriptions generated by SCA tools, rules, and standards. As we discussed in Deshapriya et al. [46], Datasets 1 and 2 were created using warnings generated by PMD and Datasets 3 and 4 were created using warnings generated by SonarQube. Results from Deshapriya et al. [46] show that the majority of the false positives are generated by a few SCA rules, and those rules are common between Datasets 1 and 2 as well as between Datasets 3 and 4. Therefore, we conducted reliability checks only in Datasets 1 and 3. To conduct reliability checking, a sample of 10% of SCA warnings from Datasets 1 and 3 were selected randomly, and two authors reviewed these samples together [67-69]. We found three data disagreements out of 115 (51 in Dataset 1 and 64 in Dataset 3) at the beginning of the reliability assessment process. The Inter-rater agreement was calculated using Cohen's Kappa between the authors [70], which was initially found to be 0.6547, indicating substantial agreement between the authors based on the categories provided by Landis and Koch [71]. After conducting the reliability check, all the authors were discussed on the samples which has disagreements and complete agreements (100%) among all the authors on the 115 samples were reached after a round of discussions [70]. Additionally, we reviewed all labels in all four datasets based on the discussion to improve overall annotation accuracy.

Moreover, all rules with no false positives were excluded from the datasets to ensure the fairness of the analysis. In Table 2, the difference between “# of rules found” and “# of rules generated false positive” represents the number of rules removed that did not produce any false positive warnings. This led to a further reduction in the size of the dataset, as shown in Table 2 under “# of samples for evaluation”. Otherwise, including only true warnings for most SCA rules would impose unnecessary overhead on performance analysis. Additionally, for simplicity of explanation, all sample datasets used in the analysis are referred to as Dataset 1, Dataset 2, Dataset 3, and Dataset 4 in other sections of this paper.

# 5. Methodology

This section outlines the methodology used to create SCA warning classifiers using LLMs. In building these classifiers, only the required slice of source code that is relevant for the SCA warnings was extracted, and the best LLMs were selected to evaluate these SCA warning classifiers. Moreover, program slicing approaches and methods for selecting the best LLM models for our study, along with LLM decoding techniques, are introduced in this section.

Program Slicing is an important step in feeding source code into an LLM for classifying SCA warnings [4, 6]. This process reduces the size of the source code fed into the LLM and ensures that the fed slices are sufficient for the LLM to classify SCA warnings [4, 6]. Program Slicing helps to speed up the classification process and minimise cost by reducing the LLM token count. After the program slicing step, prompts were designed with the program slices, warning messages and source code quality standards related to the rules that generated SCA warnings. The prompt structure is shown in Appendix G. We then feed the generated prompts for each SCA warnings to the LLMs and extract results (True Positive/False Positive) from the LLM. We then evaluate these LLMs against four manually annotated datasets [46].

## 5.1. Program Slicing

The datasets were manually reviewed and annotated to determine whether these SCA warnings are true positives or false positives. While manually annotating, we found that only a portion of the source code is relevant for deciding actionable warnings. Additionally, we identified that the Program Slicing approach should be based on the rules of SCA warnings, and warnings generated for different SCA rules require different Program Slicing techniques. Some of the warnings could be annotated by investigating only a few nearby source code lines, and we do not need to check other blocks of the source code (i.e., the issue is localised to a few nearby code lines) [73]. In contrast, some warnings require examining source code lines across different code blocks; however, there is a relationship between these code lines in terms of either data dependency or execution path, which can be identified using Data Dependency graphs or AST graphs, respectively. Therefore, irrelevant code lines exist between relevant code lines, which are unnecessary for determining the SCA warning and lead to inefficient use of the context window in the prompt [38, 74]. Consequently, we developed several program slicing approaches for each SCA rule, adapting existing Program Slicing methods, which are shown in Table 4.

```java
public class C211053{
Object foo = "x"; // CommentRequired
Integer i = (Integer)foo; //this will

}
```

A. Type 1 warnings: Only a few lines are needed to determine whether it is a true positive or a false positive.

```java
public class LocalXAConnectionFactory implements XAConnectionFactory {

    /**
    protected static class LocalXAResource implements XAResource {
        private static final Xid[] EMPTY_XID_ARRAY = {};
        private final Connection connection;
        private Xid currentXid; // @GuardedBy("this")
        private boolean originalAutoCommit; // @GuardedBy("this")

        /**
        public LocalXAResource(final Connection localTransaction) {

        private Xid checkCurrentXid() throws XAException {
            if (this.currentXid == null) {
                throw new XAException("There is no current transaction");
            }
            return currentXid;
        }

        /**
        @Override
        public synchronized void commit(final Xid xid, final boolean flag) throws XAException {

        /**
        @Override
        public synchronized void end(final Xid xid, final int flag) throws XAException {
            Objects.requireNonNull(xid, "xid");
            if (!checkCurrentXid().equals(xid)) { // LawOfDemeter
                throw new XAException("Invalid Xid: expected " + this.currentXid + ", but was " + xid);
            }
```

B. Type 2 warnings: Numerous code lines dispersed throughout the source code are necessary to determine whether it is a true positive or false positive.

*Figure 1 Code examples for warning types based on how code slices are extracted*

Figure 1.A demonstrates source code which generated an SCA warning for the rule "CommentRequired" of PMD. This SCA warning only required one code line to determine actionable SCA warnings. In such a situation, the "ISSUE_LINE_ONLY" approach was used, as shown in Table 4. In contrast, SonarQube Rule Code - "java:S1451" (Track lack of copyright and license headers) requires several code lines to determine whether the warning is a true positive or a false positive. In such scenarios, we used a few lines of code from the issue line as the slice of code. For these cases, either "ONE_LINE_UP" (the issue line and one previous line) or "ONE_LINE_DOWN" (the issue line and one following line) is used.

On the other hand, there are some warnings where the related issue lines are dispersed throughout the source code file. Therefore, identifying the relevant issue lines is not straightforward for this type of warning. The issue type – "LawOfDemeter" belongs to this kind of warning (as shown in Figure 1.B), where multiple lines of related source code are scattered throughout the source code. Therefore, following existing studies [74-77], several program slicing approaches, including "AST_BACKWARD", "AST_FORWARD", "AST_FORWARD_BACKWARD", and "DATA_DEPENDENCY", were employed when there is a code flow dependency (Data Dependency or AST) related to the classification of these kinds of warnings.

Furthermore, "AST_BACKWARD" was used to extract source code lines with AST backward dependencies, while "AST_FORWARD" was used to extract source code lines with AST forward dependencies. If source code lines with both dependencies were necessary, "AST_FORWARD_BACKWARD" was employed. Moreover, "DATA_DEPENDENCY" was utilised to extract code lines with data dependencies. AST graphs and data dependency graphs are explained in Section 2.5.

*Table 4 Source code extraction approaches*

| Program Slicing Approach | Description | Inspiration |
|---|---|---|
| ISSUE_LINE_ONLY | Extract the issue line only | Yang et al. [78] and Li et al. [75] |
| ONE_LINE_UP | Extract the issue line and the previous line from the source code | Yang et al. [78] and Li et al. [75] |
| ONE_LINE_DOWN | Extract the issue line and the following line from the source code | Yang et al. [78] and Li et al. [75] |
| AST_BACKWARD | Extract source code lines related to the backward AST graph from the issue line | Tufano et al. [76], Yerramreddy et al. [74], Li et al. [75] |
| AST_FORWARD | Extract source code lines related to the forward AST graph from the issue line | Tufano et al. [76], Yerramreddy et al. [74], Li et al. [75], and Salimi et al. [77] |
| AST_FORWARD_BACKWARD | Extract source code lines related to the backward AST graph and the forward AST graph from the issue line | Tufano et al. [76], Yerramreddy et al. [74], Li et al. [75], and Salimi et al. [77] |
| DATA_DEPENDENCY | Extract source code lines which have data dependencies on the issue line | Tufano et al. [76], Yerramreddy et al. [74], Li et al. [75], and Salimi et al. [77] |

Table 4 presents a summary of the Program Slicing approaches used in this study, along with the sources of their inspiration. The Programme Slicing methods for each PMD warning type are presented in Appendix B, while those for each SonarQube warning type are in Appendix C. A script was developed to execute the ONE_LINE_UP, ONE_LINE_DOWN, and ISSUE_LINE_ONLY slicing techniques. However, other types of Program Slicing required existing software to identify AST graphs and data dependency graphs (refer to Section 2.4). In such cases, existing studies used JOANA and Joern for program slicing [45, 46]. However, since JOANA only supports Java bytecode and older JVM versions (JDK 7 and 8) [47], Joern was employed to identify AST and data dependency graphs.

All the extracted slices were manually evaluated by the first author, and a sample of 10% from each slicing approach which is distributed over all four datasets were manually examined by the second and third authors to ensure the reliability of our Program Slicing approaches. During our manual evaluation, we did not find any missing slice that were required to determine whether the warning is true positive or false positive. Given the total agreement (100%), we did not formally compute reliability using established methods.

After extracting all the necessary line numbers, which are essential for determining whether the warning is a true positive or a false positive, the corresponding lines of source code were extracted, creating a program slice of the source code that represents the required source code to classify the warning. The resulting code snippets were used for classifying SCA warnings.

### 5.2. LLM model selection and prompting

To select LLM models for our study to evaluate, we used the LiveCodeBench leaderboard. It is based on a benchmark created by Jain et al. [79]. It is a contamination-free evaluation benchmark, where the problems/data used to evaluate the LLMs are not part of the LLMs' training process. LLMs are AI models which respond to a given query. These models are trained on a large dataset; however, it is vital to train them with data that are relevant to specific tasks [80, 81]. This study aims to classify SCA warnings into true positives or false positives using LLMs. Therefore, several factors should be satisfied by the LLM to classify SCA warnings correctly. First, the LLM should understand the patterns and structures of source code. Therefore, model fine-tuning with software source code is vital. However, since fine-tuned

LLMs with source code are already available, it is not necessary to fine-tune models in this study.

Since the LLMs are expected to behave in a deterministic manner for this study (Refer to Section 2.3), hyperparameter optimisation may not provide significant advantages. Therefore, we selected the hyperparameters based on recommendation from existing studies to improve the possibility of the LLM behaving in a deterministic manner in most situations [15, 31, 82]. Other aspects such as prompt engineering are considered in this study, as they might offer improvements in the performance of SCA warning classification when using LLMs [33, 34, 83-86]. Therefore, we used the SCoT technique to design prompts for LLMs [84]. We selected five random SCA warnings from each dataset (20 warnings altogether) and refined the prompt structure iteratively. We could reach up to 80% accuracy for the randomly selected SCA warnings. Once we fixed the prompting structure, we did not modify the structure based on the SCA rules. However, we modified the prompts for Debating Agentic pattern and iteratively optimised using ten randomly selected SCA warnings (five warnings from each rule). To evaluate our approach, we selected the top two proprietary LLMs and the top open-source LLMs from this leaderboard, as seen in Table 5 to assess the generalisability of the results of the best LLM model in the leaderboard across proprietary and open-source models. Furthermore, we accessed the leaderboard on 7th October 2025, and the LLMs have been evaluated with 579 problems within the time frame from the May 2024 to the May 2025 to create the leaderboard.

*Table 5 LLMs used in this study*

| LLM model | Comment | RQs used for |
|---|---|---|
| O4-Mini (o4-mini-2025-04-16) | Proprietary model, Same version as the source | RQ1 and RQ2 |
| Gemini-2.5-Pro (Through VertexAI) | Proprietary model, Same version as the source | RQ1 and RQ2 |
| OpenReasoning-Nemotron-32B (nvidia/OpenReasoning-Nemotron-32B) | Open-source model, Same version as the source | RQ1 and RQ2 |

To evaluate the LLMs' ability to classify SCA warnings as true positives or false positives, we designed prompts using the SCoT technique [84]. In SCoT, we structure the prompts so that the procedure to follow to classify SCA warnings is abstractly defined. A sample of the SCoT prompt is shown in Appendix G, where the role and task are defined in the first section of the prompt. In the second section of the prompt, input parameters are defined, and in the third section, the decision criteria are defined. In the decision criteria, we defined the requirements for distinguishing between a true positive and a false positive separately. Additionally, we defined the analysis steps to be followed, how to handle uncertainty, and confidence scales. Then the final section of the prompt defined the output format of the results generated by the LLM and the essential instructions to follow when analysing SCA warnings and generating predictions.

We provide several pieces of information related to SCA warnings, including the relevant program slice of the software source, the warning message, and a description of the rule associated with the SCA warning, as documented in the official SCA tool guide. The rule description is modified for more clarity or in accordance with established standards if any contradictions are found. The modified rules and their descriptions are shown in Appendix H.

We provided all the necessary information for classifying SCA warnings when prompting the LLM. Furthermore, details of the information provided to LLMs are provided in Table 6 with examples. Moreover, we used recommended hyperparameters such as temperature = 0.2 and top_p = 0.95 for similar tasks [31]. O4-mini LLM do not support temperature and top_p hyperparameters, thus, when we use O4-mini LLM, we set “high” to the reasoning effort, as the same value was used when benchmarking O4-mini via the LiveCodeBench.

*Table 6 Details of information provided with input prompts*

| **Information supplied with the input prompt** | **Description** | **Example** |
|---|---|---|
| Program Slice | Slice of the relevant source code | public void run() { // issue_line<br>final Runnable runnable = new Runnable() { |
| SCA warning message | The warning message generated by the SCA tool | Public method and constructor comments are required |
| SCA tool’s rule description | The description of the relevant rule in the official documentation of the SCA tool - this is modified from the original description to address any contradictions with established coding standards and to improve understandability | According to Clean Code book, comments are not required for classes, Constructors, Enum, fields/variables, methods. But for main methods, public non-static methods and API endpoints need comment which explain what is done by the method |

### 5.3. LLM decoding techniques and Measures for the RQs

We evaluated all the LLMs as described in Section 5.2. All the information mentioned in Table 6 are provided with the prompt for RQ1. O4-Mini, Gemini-2.5-pro, and OpenReasoning-Nemotron-32B LLM models are evaluated for answering RQ1, where we used the SCoT prompting technique. Furthermore, existing studies did not meaningfully evaluated Agentic patterns in LLMs for SCA warning classification [4, 15]. Consequently, we applied LLM agentic patterns for answering RQ2, including Self-Consistency and Debating patterns, as decoding techniques to evaluate LLMs [15]. Moreover, in RQ2, we evaluated LLMs for classifying SCA warnings using Agentic patterns with SCoT and measured their performance to determine whether Agentic patterns improve warning classification. The procedure we followed for the Debating pattern is shown in Figure 2. We set the temperature to 0.7 for both the Prosecutor Agent and the Defender Agents, and to 0.4 for the Critic Agent [15, 31, 82]. Additionally, we set the temperature to 0.2 for the Arbiter Agent. Furthermore, we used 0.95 as top_p for every agent [15, 31, 82]. Moreover, the procedure we followed for the Self-Consistency pattern is shown in Figure 3, where we used 0.7 for the temperature hyperparameter and 0.95 for top_p following the original paper for the Self-Consistency [82]. Additionally, because O4-mini LLM lacks support for temperature and top_p hyperparameters, we consistently set reasoning effort to high, matching the value used during the O4-mini benchmarking in LiveCodeBench. Moreover, results for each research question are reported in MCC, F1-Score, BA, and time taken to classify a SCA warning to understand the usability of each LLM in terms of the purpose of building classifiers as described in the Section 2.2. Furthermore, the methodologies followed in answering each research questions are summarised in Table 7.

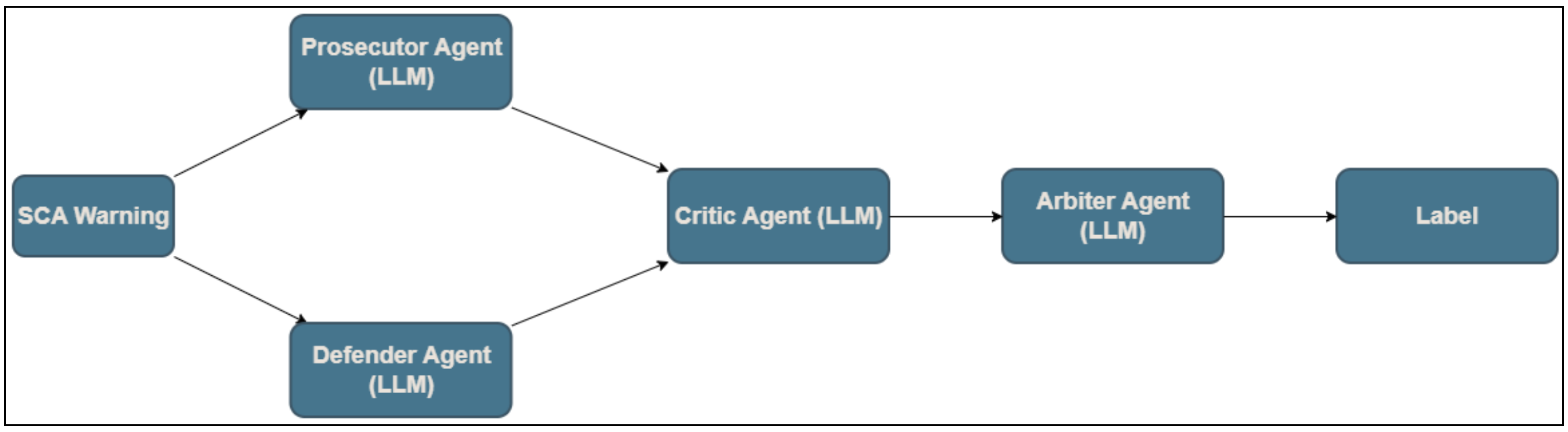

*Figure 2 Debating Agentic pattern with Structured CoT request for RQ2*

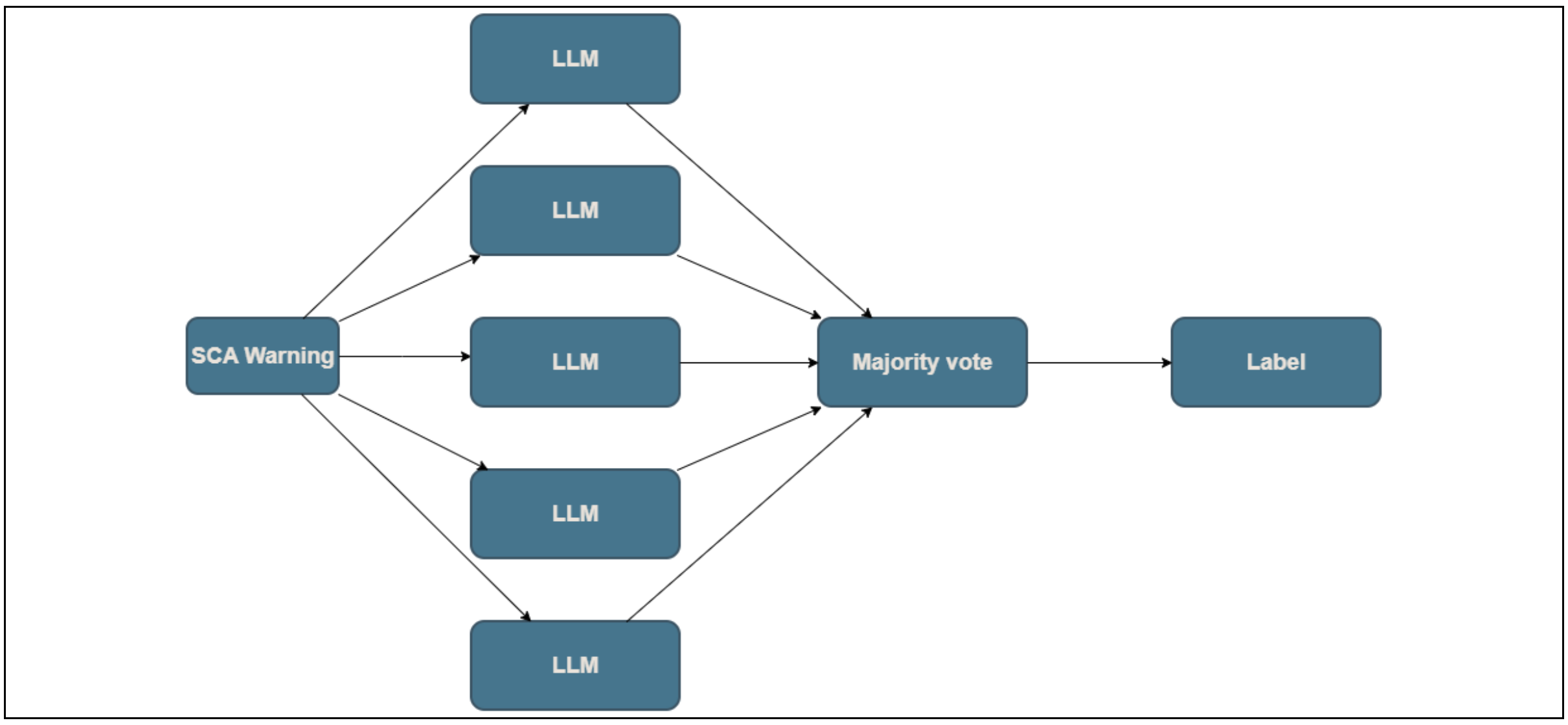

*Figure 3 Self-Consistency Agentic pattern with Structured CoT request for RQ1*

*Table 7 Methodology followed to prepare data to answer RQs*

| Research Question | Methodology |
|---|---|
| RQ1:What is the performance of pre-trained models in classifying SCA warnings using established coding standards? | Each LLM model was separately evaluated with four datasets. The instances that false positives were correctly/incorrectly identified by LLMs using SCoT prompting approach are collected individually. Time taken to analyse SCA warnings were also recorded and cost was also calculated in relevant scenarios. These results were reported and analysed in Section 6.1. Moreover, Cochran's Q test was conducted among LLMs for each dataset separately to check if the performance difference between each LLM is significant. Additionally, Pairwise McNemar tests were conducted to identify which LLMs produced significantly better performance among the datasets. |
| RQ2:How do the Agentic Patterns (Debating and Self-Consistency) elevate the performance of the best pre-trained models in classifying warnings generated by SCA tools? | We identified the rules which generated the most incorrect results in RQ1. Of the rules, most incorrect results generated from each SCA tool were selected and examined further in RQ2. Similar to RQ1, LLMs were evaluated using Agentic patterns such as Self-Consistency and Debating. The performance of each LLM for each dataset, time taken to classify SCA warnings and costs are reported and analysed in Section 6.2. Moreover, Cochran's Q test was conducted among LLMs for each dataset separately to check if the performance difference between each LLM is significant. Additionally, Pairwise McNemar tests were conducted to identify which LLMs produced significantly better performance among datasets. Furthermore, Cochran's Q tests were conducted among different runs in the Self-consistency approach to understand the impact of higher temperature values for LLMs' performance variability. |

# 6. Results

## 6.1. RQ1: What is the performance of pre-trained models in classifying SCA warnings using established coding standards?

Our four datasets, discussed in Section 4, were used to evaluate the selected LLM models from Section 5.2. These LLMs classified the SCA warnings in our datasets independently of the manually assigned labels, and later, the predicted labels were compared with the manually assigned labels to assess the performance of our approach. We used the SCoT methodology as the prompting technique. The results for the top 5[7] rules which generated the most SCA warnings are shown in Table 8, Table 9, Table 10 and Table 11 for Datasets 1, 2, 3 and 4, respectively. The complete results for each rule and dataset are shown in Appendix D, Appendix E, and Appendix F for Datasets 1, 2, and 4, respectively. Moreover, Table 10 shows the complete results for Dataset 3.

*Table 8 Correct and incorrect number of warnings for the top five rules in Dataset1*

| **LLM** | **Rule (Correct/Incorrect)** | | | | |
|---|---|---|---|---|---|
| | **CommentRequired** | **UseUtilityClass** | **LawOfDemeter** | **SingularField** | **UnusedAssignment** |
| **OpenReasoning** | 102/3 | 12/0 | 9/1 | 2/0 | 2/0 |
| **Gemini** | 97/8 | 12/0 | 9/1 | 1/1 | 1/1 |
| **O4_mini** | 101/4 | 12/0 | 9/1 | 2/0 | 2/0 |

*Table 9 Correct and incorrect number of warnings for the top five rules in Dataset2*

| **LLM** | **Rule (Correct/Incorrect)** | | | | |
|---|---|---|---|---|---|
| | **CommentRequired** | **LawOfDemeter** | **JUnitTestsShouldIncludeAssert** | **CloseResource** | **UselessParentheses** |
| **OpenReasoning** | 38/14 | 22/23 | 2/2 | 2/0 | 1/2 |
| **Gemini** | 39/13 | 26/19 | 3/1 | 2/0 | 1/2 |
| **O4_mini** | 41/11 | 21/24 | 2/3 | 2/0 | 1/2 |

*Table 10 Correct and incorrect number of warnings for each rule in Dataset3*

| **LLM** | **Rule (Correct/Incorrect)** | | | | |
|---|---|---|---|---|---|
| | **Source code should be indented consistently** | **Magic numbers should not be used** | **equals(Object obj) should test the argument's type** | **Lazy initialization of "static" fields should be "synchronized"** | **Classes should not have too many fields** |
| **OpenReasoning** | 104/1 | 28/14 | 1/0 | 1/0 | 0/1 |
| **Gemini** | 103/2 | 28/14 | 1/0 | 1/0 | 1/0 |
| **O4_mini** | 103/2 | 27/15 | 1/0 | 1/0 | 0/1 |

*Table 11 Correct and incorrect number of warnings for the top five rules in Dataset4*

| **LLM** | **Rule (Correct/Incorrect)** | | | | |
|---|---|---|---|---|---|
| | **Source code should be indented consistently** | **Magic numbers should not be used** | **Track lack of copyright and license headers** | **Member variable visibility should be specified** | **Arrays and lists should not be copied using loops** |
| **OpenReasoning** | 223/29 | 14/7 | 2/7 | 1/5 | 0/1 |
| **Gemini** | 245/7 | 11/10 | 1/8 | 1/5 | 1/0 |
| **O4_mini** | 248/4 | 7/14 | 8/1 | 0/6 | 1/0 |

The performance of each LLM for each dataset is shown in Figure 4 in terms of performance metrics, including Precision, Recall, MCC, F1-Score, BA, and average time taken to analyse a warning. Figure 4 show that Dataset 1 achieved the best performance across all metrics. Moreover, the OpenReasoning model performed best across most datasets and most performance metrics. However, Gemini was the best LLM for Dataset 2 in terms of Precision, Recall, MCC, F1-Score, and BA, and O4-mini was best for Dataset 4 in terms of Precision,

[7] "top 5" means the results for the five rules that generated the most warnings.

MCC and F1-Score. Moreover, OpenReasoning models took significantly (5.8-7.1 times slower than the median) longer to analyse warnings than other LLMs for all the datasets. As described in Section 5, we deployed the OpenReasoning model on a locally hosted single-GPU server. Therefore, the performance of the OpenReasoning models in terms of time to produce results is noteworthy.

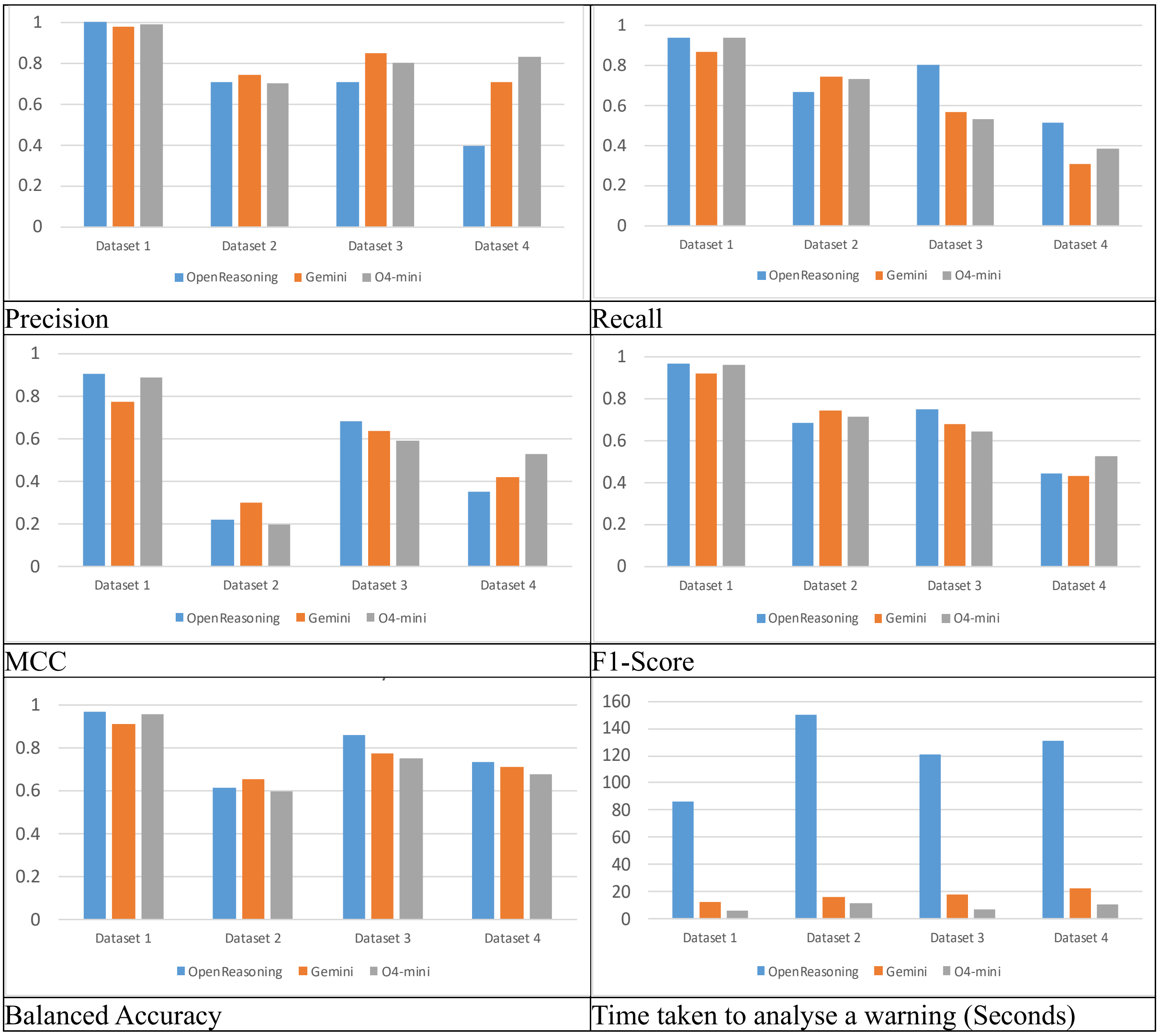

*Figure 4 Performance of SCA warning classification using SCoT for all the warning types*

*Table 12 Cochran's Q test results for all datasets among LLMs (Significant values were bolded)*

| **Dataset** | **Accuracy** | | | **Q statistic** | **p value** |
|---|---|---|---|---|---|
| | **OpenReasoning** | **Gemini** | **O4-mini** | | |
| 1 | 0.9496 | 0.8921 | 0.9496 | 10.6667 | **0.0048** |
| 2 | 0.6068 | 0.6752 | 0.6325 | 2.5128 | 0.2847 |
| 3 | 0.8933 | 0.8933 | 0.8800 | 0.4000 | 0.8187 |
| 4 | 0.8322 | 0.8926 | 0.9094 | 14.3934 | **0.0007** |
| All | 0.8310 | 0.8565 | 0.8651 | 7.0909 | **0.0289** |

Each SCA warning in each dataset was fed into the LLM to predict the label, and the predicted label was compared with the actual label. The labels are binary, therefore, to compare the results, Cochran's Q test was conducted [87, 88], and the results are shown in Table 12. First, we conducted these statistical tests on each dataset, and then on all SCA warnings across all datasets. Therefore, the last row of the Table 12 shows the statistical test results for the collection of all SCA warnings across all datasets. Table 12 demonstrates the accuracy of each LLM for each dataset, Q statistic values, and the p-values. The significant p values ($p < 0.05$) are bolded. According to Table 12, for Datasets 1, 4 and All, at least one LLM produced significantly better performance over other LLMs.

*Table 13 Pairwise McNemar test results among LLMs with Holm-Bonferroni Correction(Significant values were bolded)*

| Dataset | Gemini vs O4-Mini | | OpenReasoning vs Gemini | | OpenReasoning vs O4-Mini | |
|---|---|---|---|---|---|---|
| | P Value | Adjusted_alpha | P Value | Adjusted_alpha | P Value | Adjusted_alpha |
| 1 | **0.013328** | 0.016667 | 0.043308 | 0.025000 | 0.617075 | 0.050000 |
| 4 | 0.404248 | 0.050000 | **0.014138** | 0.025000 | **0.002066** | 0.016667 |
| All | 0.531971 | 0.050000 | 0.101876 | 0.025000 | **0.016489** | 0.016667 |

To identify the exact LLMs which generated significantly better performance, Pairwise McNemar tests with Holm-Bonferroni Correction were executed [89]. These results are shown in Table 13 where O4-mini is seen to be significantly better than Gemini on Dataset 1. Additionally, Gemini and O4-mini are significantly better than OpenReasoning LLM for Dataset 4, and O4-mini is better than OpenReasoning for all the datasets collectively. Consequently, the LLMs show no significant performance differences on Datasets 2 and 3; in other scenarios, O4-mini performs best.

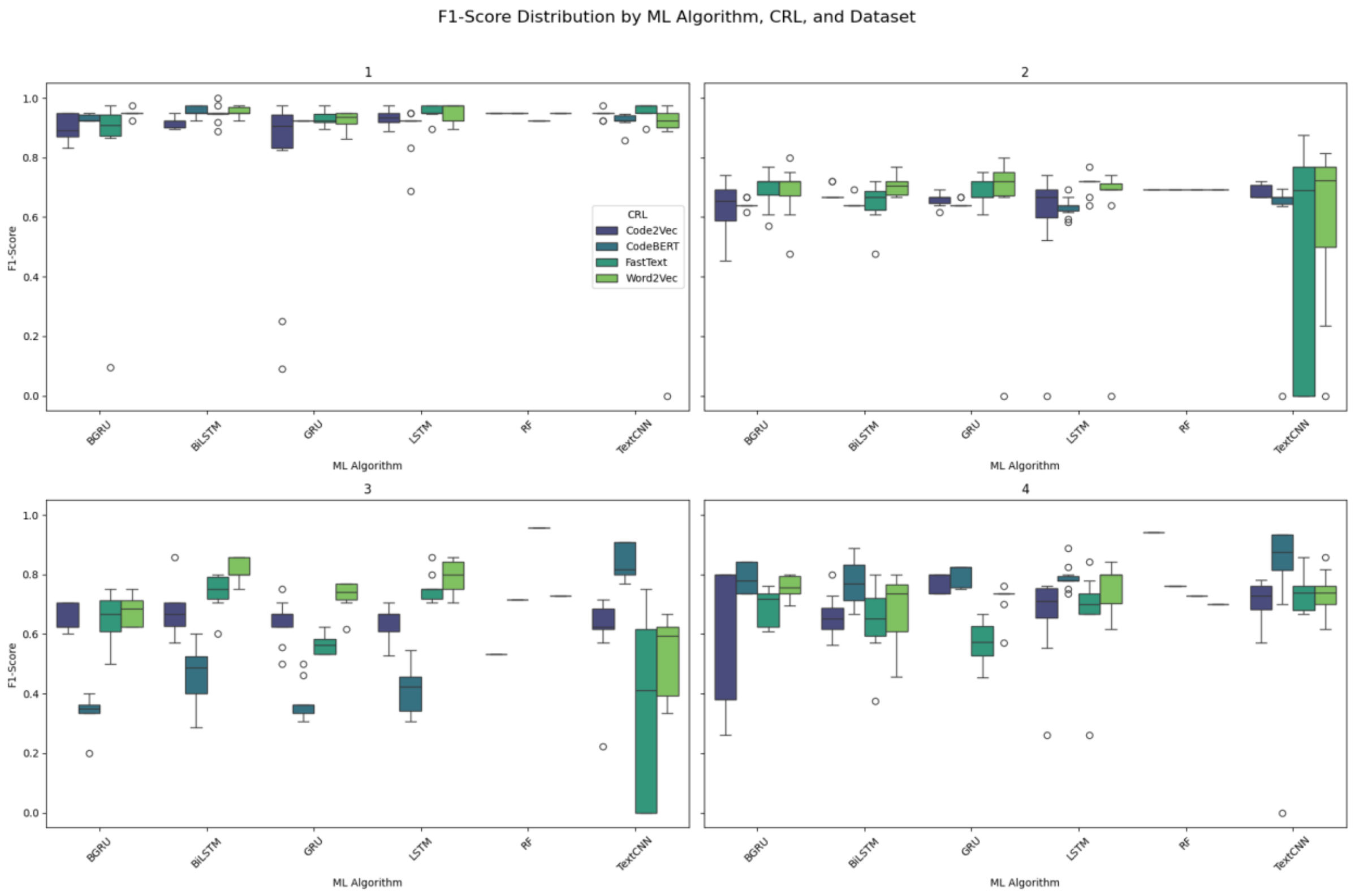

*Figure 5 Box plots for ML models trained and evaluated on the same datasets*

We selected few Machine Learning (ML) approaches as baseline for our study as ML approaches were popular in academia for SCA warning classification before LLMs were introduced [74, 78, 90, 91]. We used four Code Representation Learning (CRL) models (Code2Vec, CodeBERT, FastText and Word2Vec) to generate numerical representations of source code program slices to ML algorithms. We utilised six ML algorithms such as BGRU, BiLSTM, GRU, LSTM, Random Forest, and TextCNN). Figure 5 shows the results for ML approaches trained on the same datasets from our previous study [92]. Datasets were trained with multiple ML models 10 times each and graphs in Figure 5 show boxplots for F1-Score values for each iteration. According to the Figure 5, the best ML model is varies based on the dataset. However, some ML models performed better than LLM approaches evaluated on the same datasets shows that LLM models produce lower results, mainly for Datasets 2-4.

### 6.2. RQ2: How do the Agentic Patterns (Debating and Self-Consistency) elevate the performance of the best pre-trained models in classifying warnings generated by SCA tools?

RQ1 focused on identifying the performance of the three best frontier LLM models in classifying SCA warnings. The results for the top 5 rules which generated the highest number of warnings are shown in the Table 8 to Table 11. According to these results, the "LawOfDemeter" rule produced the most incorrect results in PMD, while the "Magic numbers should not be used" rule generated the most incorrect results for SonarQube. Therefore, for RQ2, further investigations were conducted only for warnings generated by these rules.

In RQ2, we examined the use of Agentic patterns proposed by Gullí [15]. We used Self-Consistency and Debating patterns to improve the warning classification task, and the results are compared with SCoT in Figure 6. According to Figure 6, results generated for our analysis show that although the use of Agentic patterns yielded lower performance, the O4-mini LLM model achieved higher performance in some scenarios, such as classifying warnings generated by the SonarQube rule "Magic numbers should not be used," as measured by Precision, MCC, F1-Score, and BA. Furthermore, according to Figure 6, the use of Agentic patterns required more time to analyse a warning, as it needed multiple LLM calls for each SCA warning. Agentic patterns seem unnecessary for most scenarios, and further investigation is required in order to identify practical applications where they are useful.

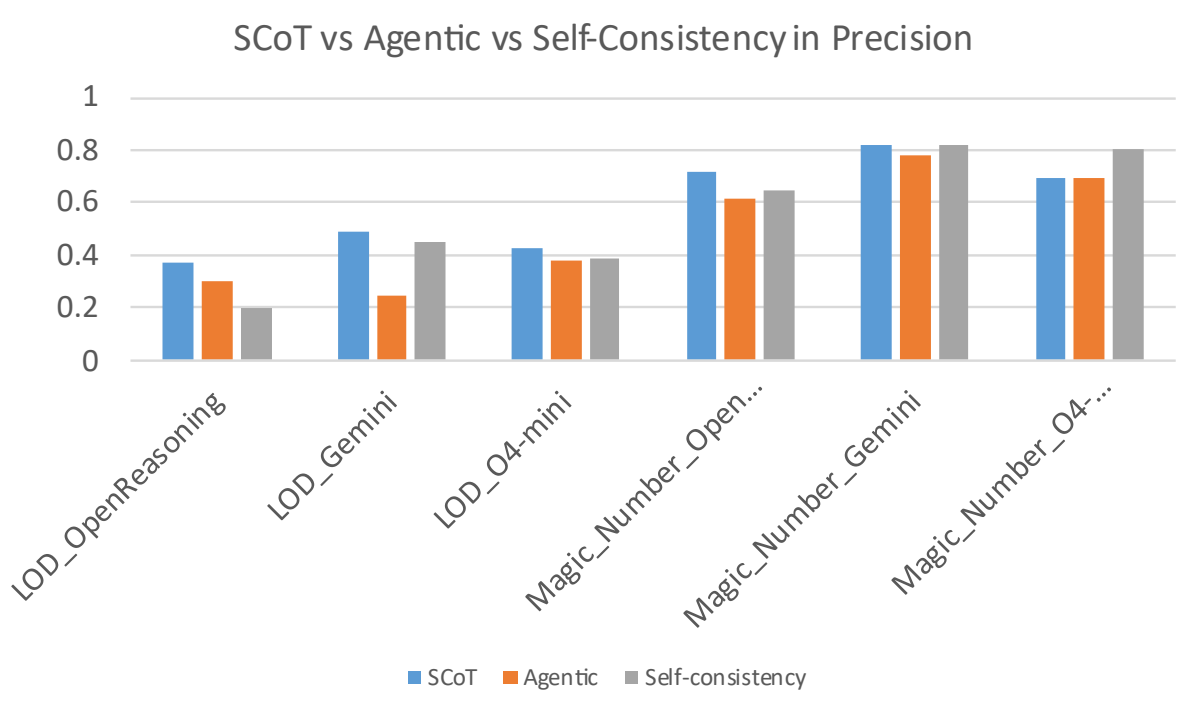

Precision

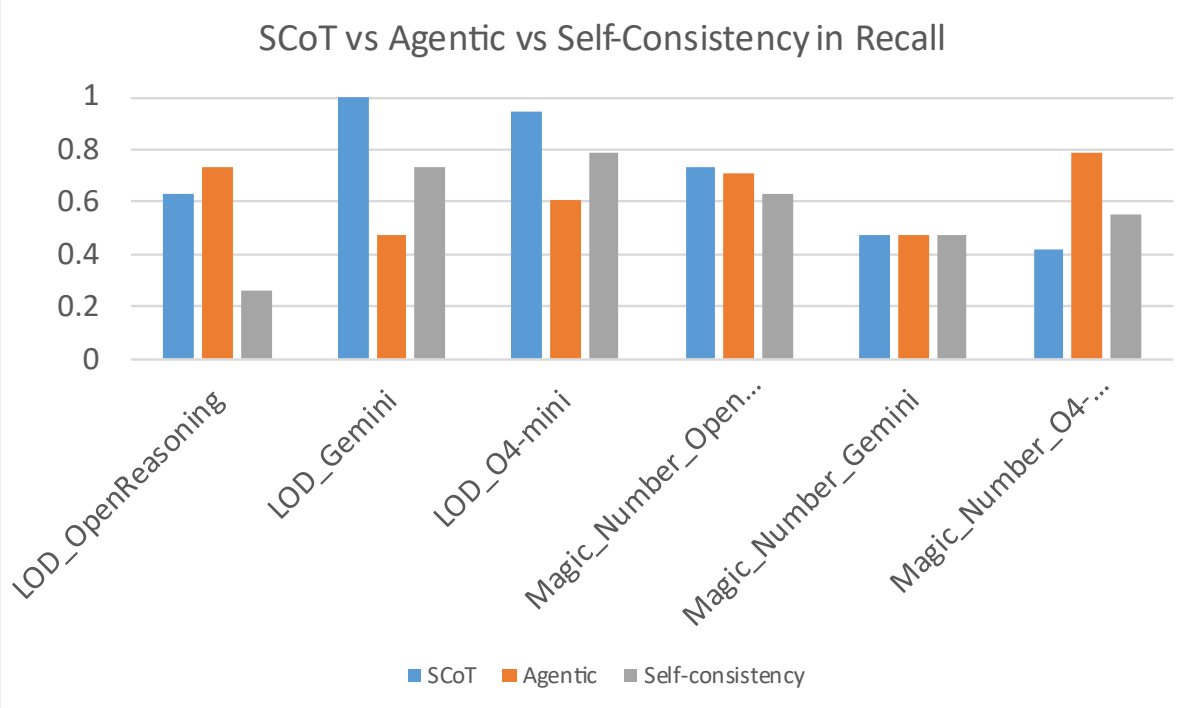

Recall

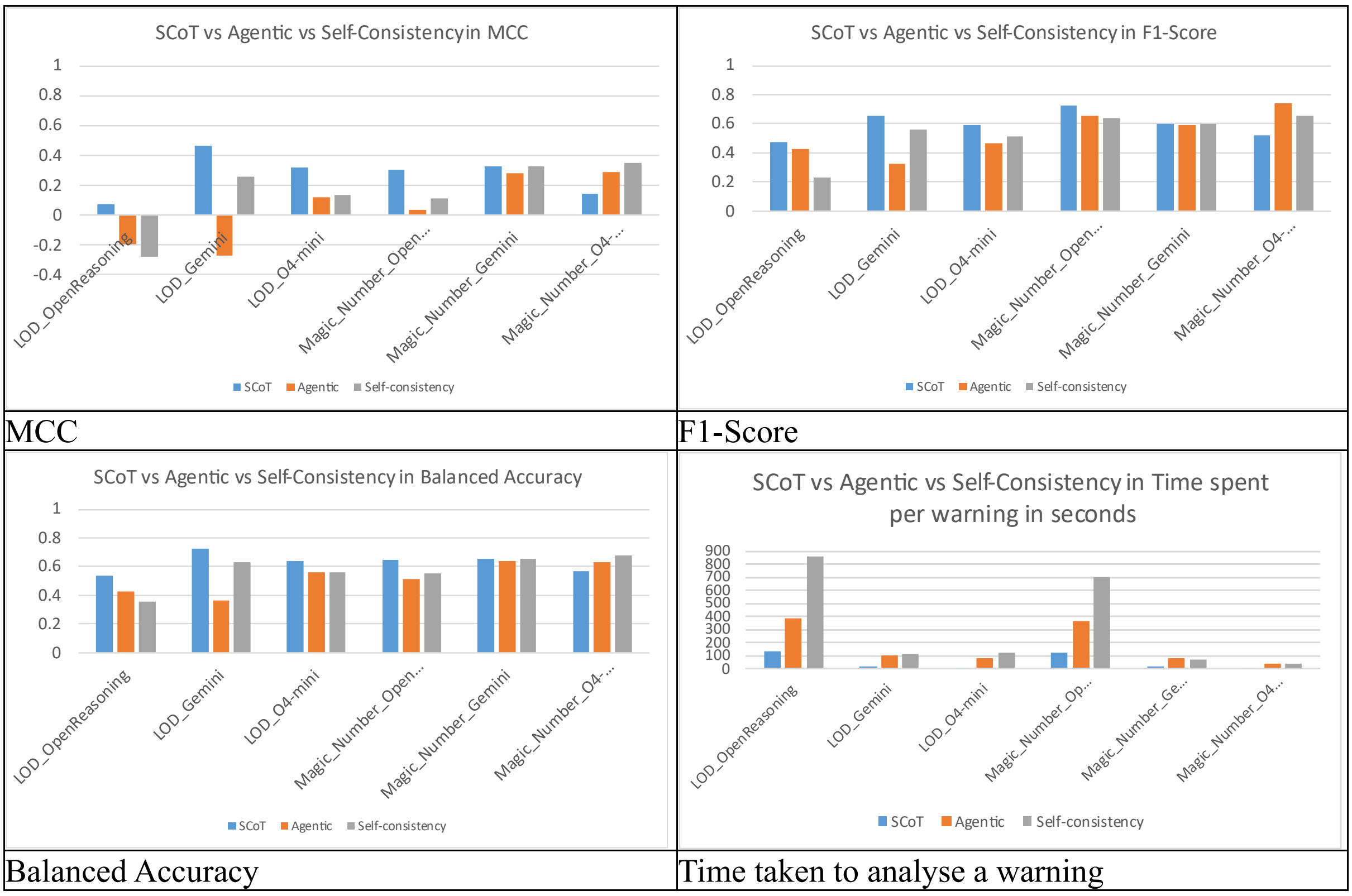

*Figure 6 Performance of SCoT, Agentic AI and Self-consistency implementations for selected warning types*

*Table 14 Cochran's Q test results for all issueType_LLM among different prompting techniques*

| **IssueType_LLM** | **Accuracy** | | | **Q statistic** | **P value** |
|---|---|---|---|---|---|
| | **SCoT** | **Debating** | **Self-Consistency** | | |
| LOD_OpenReasoning | 0.5091 | 0.3273 | 0.3818 | 4.6471 | 0.0979 |
| LOD_Gemini | 0.6364 | 0.3273 | 0.6000 | 15.2353 | **0.0005** |
| LOD_O4-mini | 0.5455 | 0.5273 | 0.4909 | 0.5185 | 0.7716 |
| Magic_OpenReasoning | 0.6667 | 0.5556 | 0.5714 | 3.3077 | 0.1913 |
| Magic_Gemini | 0.6190 | 0.6032 | 0.6190 | 0.1176 | 0.9429 |
| Magic_04-mini | 0.5397 | 0.6667 | 0.6508 | 4.0714 | 0.1306 |
| OpenReasoning (LOD and Magic) | 0.5932 | 0.4492 | 0.4831 | 7.9000 | **0.0193** |
| Gemini (LOD and Magic) | 0.6271 | 0.4746 | 0.6102 | 11.451 | **0.0033** |
| O4-mini (LOD and Magic) | 0.5424 | 0.6017 | 0.5763 | 1.3455 | 0.5103 |

Cochran's Q tests were conducted for all issueType_LLM combinations among different prompting techniques, and the results are shown in Table 14 [87, 88]. Table 14 shows the accuracy of SCA warning classification in each prompting technique and Q statistic values, and p values for each issueType_LLM combination. The significant p values ($p < 0.05$) are bolded in Table 14, where LOD_Gemini and OpenReasoning (for both issue types collectively) and Gemini (for both issue types collectively) have at least one significant prompting technique. However, for other IssueType_LLM combinations, all the prompting techniques perform without any significant performance difference.

*Table 15 Pairwise McNemar test results among different prompting techniques with Holm-Bonferroni Correction*

| **IssueType_LLM** | **SCoT vs Debating** | | **Debating vs Self-Consistency** | | **SCoT vs Self-Consistency** | |
|---|---|---|---|---|---|---|
| | **P Value** | **Adjusted_alpha** | **P Value** | **Adjusted_alpha** | **P Value** | **Adjusted_alpha** |
| LOD_Gemini | **0.002076** | 0.016667 | **0.009330** | 0.025000 | 0.772830 | 0.050000 |
| OpenReasoning | 0.017073 | 0.016667 | 0.651077 | 0.050000 | 0.031141 | 0.025000 |
| Gemini | **0.008712** | 0.016667 | **0.017706** | 0.025000 | 0.823063 | 0.050000 |

To identify the exact prompting technique which generated significantly better performance, Pairwise McNemar tests with Holm-Bonferroni Correction were conducted for the IssueType_LLM [89]. The results are shown in Table 15, where for LOD_Gemini and Gemini (LOD and Magic), both SCoT and Self-Consistency perform significantly better than Debating. OpenReasoning (LOD and Magic) shows a sign of the existence of a significant prompting technique, according to Cochran's Q test. However, after applying the Holm-Bonferroni correction for the pairwise McNemar test, there is no prompting technique that performs significantly better than the other techniques.

On the other hand, we ran the same set of SCA warnings five times to conduct a Self-Consistency approach. Given the use of 0.7 as the temperature value for the LLM calls, we used the results from individual LLM calls made for the Self-Consistency approach to validate the potential impact of the higher temperature value on the overall performance variation of the LLMs in different runs. We conducted Cochran's Q tests for each dataset and each LLM. The results are shown in Table 16, where p values for every scenario are higher than 0.05, indicating that there are no significant performance differences between runs.

*Table 16 Cochran's Q test results for LLM performance among different runs in the Self-consistency approach*

| | **Dataset 1 (Q Statistic/P Value)** | **Dataset 2 (Q Statistic/P Value)** | **Dataset 3 (Q Statistic/P Value)** | **Dataset 4 (Q Statistic/P Value)** |
|---|---|---|---|---|
| Gemini | 2.285/0.683 | 7.081/0.131 | 2.800/0.591 | 5.846/0.210 |
| O4-Mini | 4.666/0.323 | 3.593/0.463 | 2.086/0.719 | 7.076/0.131 |
| OpenReasoning | 6.352/0.174 | 5.090/0.278 | 5.492/0.240 | 1.166/0.883 |

Moreover, we conducted hyper-parameter tuning for a small set of temperature (T) values such as 0.0, 0.7, 1.0 and 2.0 for self-consistency for the OpenReasoning model. During the experiments, T = 2.0 could not generate final answers within the given maximum number of tokens allowed (16,384). A summary of results for other temperature values is shown in Table 17, where it can be seen that T = 0.7 is the best temperature value for our study overall. While T = 1.0 also performed well, T = 0.7 performed best.

*Table 17 Hyperparameter tuning results for each dataset for multiple temperature values*

| Dataset | T = 0.0 | T = 0.7 | T = 1.0 |
|---|---|---|---|
| Dataset 1 | 5/10 | 6/10 | **7/10** |
| Dataset 2 | 21/45 | **26/45** | 20/45 |
| Dataset 3 | **18/42** | 16/42 | 15/42 |
| Dataset 4 | 10/21 | **11/21** | **11/21** |

We used three LLMs for our analysis and we paid O4-mini through the OpenAI platform. Moreover, we used the Gemini 2.5 pro Api through a free Vertex AI account[8], where we did not pay for the api calls. Additionally, our OpenReasoning-Nemotron-32B model was deployed in our university servers, where it is difficult to estimate the cost of LLM calls. Therefore, we conducted a cost analysis for the O4-mini model to get an understanding of the cost of each approach we experimented with in this paper. O4-mini LLM model cost 0.009134 USD on average for SCoT where only one LLM call was made for one SCA warning. Additionally, the cost to classify one SCA warning using Self-consistence was 0.045668 USD and 0.036535 USD for using Debating. SCoT generates only one LLM request per SCA warning, whereas five are generated for Self-Consistency and four for the Debating approach. On the other hand, Self-Consistency and Debating approaches did not generate significant performance improvements, according to Figure 6. According to our analysis, investing more in Agentic patterns is not cost effective.

# 7. Discussion and implications

## 7.1. RQ1: What is the performance of pre-trained models in classifying SCA warnings using established coding standards?

RQ1 examined the performance of the best pre-trained LLMs for SCA warning classification. To classify SCA warnings, we used relevant source code slices, warning messages generated by the SCA tool and the rule descriptions applicable to the generated warning type. Moreover, we checked the rule descriptions against the established coding standards and, if there was any contradiction, modified the descriptions accordingly. All three LLMs achieved good results on Dataset 1 across all performance metrics, including MCC, F1-Score, BA, and time taken to analyse a warning. However, all the LLMs generated lower results for the other datasets. Datasets 1 and 2 consist of warnings generated by PMD, while Datasets 3 and 4 consist of warnings generated by SonarQube. Furthermore, even within Datasets 1 and 2 (and Datasets 3 and 4), the number of warnings per SCA rule varies. Therefore, the choice of LLMs for SCA warning classification is based on the performance metrics and designated goals (Refer to Section 2.2). According to Section 3, most of the existing studies do not incorporate necessary contextual information, such as SCA rule description and established standards for SCA warnings classification [2, 4, 25]. Without the contextual information, LLM will use their internal knowledge, which may be incorrect or outdated. Additionally, the standards for defining warnings are true positives or false positives, depending on the individual preferences. Therefore, it is vital for the SCA warning classifier to support defining individual preferences instead of using internal knowledge of LLMs. Reviewing existing studies summarised in Section 3 [4, 6, 10], none of these studies attempted to provide defined code standards (possibly different based on the software organisation or the team) when prompting LLMs so that the warning classification tasks are performed based on the defined standards. We observed in our previous study that warning classification is based on defined standards [46]. Therefore, it is vital to provide coding standards with the LLM prompts. However, existing studies did not address this requirement. Moreover, existing studies primarily rely on publicly available datasets to evaluate LLMs in SCA warning classification. However, the threat of using publicly available datasets for this designated purpose is that LLMs might have incorporated them into their training, and LLMs might memorise and retrieve labels for SCA warnings from their internal knowledge rather than predicting them. However, when these LLMs are applied to real-world datasets in a real software project, they produce lower results. This aspect is evaluated by multiple existing studies [2, 4, 6]. However, none of them experimented on multiple datasets with different false positive percentages. Only Chen et al. [2, 4] tried to identify the real threat of the usage of publicly available datasets to evaluate LLMs in classifying SCA warnings.

[8] https://cloud.google.com/products/gemini-enterprise-agent-platform

Furthermore, none of the existing studies conducted any experiments for both ML models and LLMs to compare the performance among ML and LLM technologies. We experimented with several ML models on the same dataset that is used to evaluate LLMs in this study. According to the results, ML models perform better than LLM models. Even simpler models such as Random Forest performed better than LLMs. These Random Forest models can train on commodity hardware without any costly GPUs and produces results faster than LLMs tend to respond. Therefore, we can conclude that ML models are better than LLM models in classifying SCA warnings at present.

In this study, we address gaps identified in existing studies satisfactorily. We used our own dataset [16, 17], which was not used to train LLMs, and source code datasets used to create warning datasets are align with real world code. Thus, our results show the genuine performance of the real applications in the practitioner community. Even though existing studies show that LLMs generate near-perfect results [2, 4, 6, 25], the reality is different when unseen data is used for modelling. A summary of SCA warning classification performance is shown in Appendix A, where our approach achieved better results even on real-world datasets compared to existing studies that used them to evaluate their approaches. Moreover, LLMs generate a description of their decision, which makes SCA warning classification with LLMs more explainable than using other ML approaches blindly.

### 7.2. RQ2: How do the Agentic Patterns (Debating and Self-Consistency) elevate the performance of the pre-trained best models in classifying warnings generated by SCA tools?

The performance of the LLMs for our datasets was worse than existing similar studies (Refer Section 3). Therefore, we sought to improve performance by applying some Agentic patterns to our approach. Consequently, we selected a warning type from each SCA tool and applied these Agentic patterns only to those warning types. The results show that the performance did not improve, even using Agentic patterns, except for warnings generated by the “Magic numbers should not be used” rule in SonarQube with O4-mini LLM. There can be two possible reasons for this. The first reason could be that we use a higher temperature (0.7) for Agentic patterns as recommended in the original study to increase the creativity of the generated results [15, 82]. Although, this temperature value is not recommended for tasks such as code generation and understanding related tasks [31], existing studies generated near-perfect results for publicly available benchmark datasets in classifying SCA warnings. Since LLMs may have been trained on these datasets, these existing studies generated near-perfect results while LLMs may generate worse results for unseen datasets. However, the O4-mini LLM does not support temperature or top_p hyperparameters. Therefore, we used the "reasoning effort" hyperparameter. We used "high" for both RQ1 and RQ2. Moreover, O4-mini reduced the performance for LOD when using Agentic patterns and improved the performance for the "Magic numbers should not be used" rule. Therefore, the performance of LLMs may be dependent on the rule type as well. However, for most LLMs, using Agentic patterns in our datasets is unnecessary.

Additionally, we use SCoT when designing prompts for LLMs. Therefore, there is little room for LLMs' performance to improve or deteriorate, even though they use these Agentic patterns. Comparing our study with existing studies reveals that only a few have used a self-consistency-agentic approach. However, none of the existing studies used a Debating approach. In our study, we showed that using Agentic patterns for SCA warning classification is not beneficial, as it incurs a higher cost due to multiple LLM API calls, takes longer to process each warning, and does not significantly improve performance. Additionally, when Self-Consistency was added, the performance of SCA warning classification deteriorated in most scenarios.

### 7.3. Implications for SCA tool developers

SCA tool developers need to reduce false positive warnings generated by their tools. For that, they can use LLMs to identify false-positive warnings generated by their SCA tools and common patterns that cause them to issue false positives. Consequently, by identifying patterns, tool developers can improve SCA tool performance. On the other hand, the goal of SCA tool developers is to identify all issues in source code, even if they generate false positive warnings. However, users of the SCA tool (the practitioner community) may need to identify as many warnings as possible while minimising false positives. Therefore, it is vital to introduce a sensitivity parameter that allows SCA tool users to adjust based on their needs. Additionally, SCA tool developers can publish statistics on false positive rates for each rule. Therefore, users can deactivate rules that generate a high number of false positive warnings if those warnings are not critical to their software projects.

### 7.4. Implications for the Practitioner community

The purpose of creating the SCA warning classifier is discussed in Section 2.2, and according to that, there are multiple goals. Moreover, goals vary depending on individual preferences and use case. For instance, one might need to reduce false positives as much as possible, even though the classifier removes some genuine warnings as well. Another situation might be to identify the maximum number of genuine warnings, while minimising false warnings as much as possible. One might need to optimise overall model performance, while another may need to consider the trade-off between model performance and the time required to process warnings. Therefore, the practitioner community should use appropriate performance metrics to evaluate LLMs against their requirements, rather than rely solely on a few popular metrics such as accuracy and F1-Score. Furthermore, practitioner communities can host open-source LLMs on their own servers or use proprietary services such as ChatGPT, Gemini, and Claude. According to results generated from our study (refer to Section 6), proprietary services are faster than self-hosted services, depending on the performance of the hosted hardware. However, there is a potential risk to data privacy, and the practitioner community should also consider this.

Additionally, it is useful to experiment with different LLMs, structured prompts and prompting techniques, decoding techniques and hyperparameters to find the best combination for individual use cases, as most cutting-edge technologies/approaches may not be the best for some use cases. The results are summarised in Section 6 show that the performance of the classifier depends on the characteristics of the dataset, the purpose of building the classifier (refer to Section 2), and the decoding technique used. Therefore, it is necessary to experiment with the variables mentioned earlier to find the best combination for each use case and dataset.

Coding standards are not universal. Standards can depend on the software engineering company, client organisation, project team and the programming language and framework used. We identified some defined standards in our previous study; however, those should be tailored to different use cases. Additionally, standards evolve over time. Therefore, coding standards should be configurable in SCA warning classifiers to cater for this requirement. Moreover, the practitioner community can use the description generated by LLM to determine whether the SCA warning is a true positive or a false positive, rather than relying solely on the label generated by the classifier. Using the description, the developer can understand whether the SCA warning is a true positive or a false positive.

The results show that the model performance depends on the LLM, dataset and performance metric. Therefore, practitioners should select performance metrics based on their goals (Refer to Section 2.2). Furthermore, practitioners should evaluate LLMs on their private datasets as their performance is heavily dependent on the characteristics of the dataset. Moreover, LLMs provide description with the final answer (whether a SCA warning is a true positive or false positive). The description provides good understanding of the reason for the classification, providing model explainability. Results from this study shows that ML models provide better performance. However, explainability of ML models are lower, requiring separate consideration (e.g., perhaps via the use of SHAP or LIME). Therefore, the explainability aspect of a model should also be considered when selecting technologies to create warning classification models. On the other hand, Agentic patterns did not improve the performance for our datasets. Therefore, practitioners should not assume that Agentic patterns may always improve the performance outcomes. The final decision should be made based on experiments conducted on their datasets.

### 7.5. Implications for academia

Most existing studies evaluating LLMs on public datasets/benchmarks suggest that LLMs perform perfectly in SCA warning classification. However, the LLM may have been trained on a given publicly available dataset/benchmark, giving LLMs the opportunity to memorise the label for a specific SCA warning. Consequently, when the same approach is applied to real-world use cases, LLMs may perform worse. Therefore, academia should create benchmarks to evaluate SCA warning classification using LLMs which do not incorporate the LLM's training data, similar to LiveCodeBench. Furthermore, the use of high-quality evaluation datasets that mirror the characteristics of real-world use cases is vital. Otherwise, the evaluation results are not compatible with the real-world use cases. Conversely, academia can report evaluation results across datasets with varying characteristics, such as different false positive rates and SCA tools. Additionally, it is vital for academia to report their experiment results in multiple performance metrics, so that the community can choose SCA warning classification approaches suitable for their use cases, reflecting their goal of warning classification (refer to Section 2.2). Moreover, most existing studies only experiment with specific LLMs and prompting techniques to classify SCA warnings. However, academia can focus on hyperparameter optimisation and model fine-tuning, specialising in warning classification, leading to improved model performance. Additionally, to speed up experiments, academia can first test their approaches on small datasets to evaluate different technologies and methodologies, thereby accelerating the refinement of their methods. This will speed up similar studies, as it takes academia a long time to experiment with LLMs across large datasets. Consequently, we recommend our datasets to the academic community for their experiments, as they are relatively small yet mirror real-world scenarios. These datasets were created by analysing publicly available source code datasets with PMD and SonarQube. Moreover, results show that LLMs produce lower performance for unseen datasets. In this study, we did not evaluate the reason for LLMs producing lower results. We suggest that future research should focus on conducting deeper qualitative analyses in this regard.

# 8. Threats to validity

This section discusses the potential threats to the validity of this study following some existing studies [2, 24, 25, 93].

**Internal threats to validity**

As pointed out by Croft et al. [94], AI/ML model evaluation depends on the quality of the dataset used. If the dataset is incorrectly classified, the evaluation results differ from the actual performance statistics. Therefore, it is vital to use a high-quality dataset. According to Lin et al. [95], the best approach to creating such a high-quality database is manual annotation rather than automated methods. On the other hand, publicly available datasets might have been used to train existing frontier LLMs. Consequently, using publicly available datasets to evaluate LLMs for SCA warning classification can be ineffective. To mitigate these issues, we created four datasets and annotated them manually, as explained in Section 4. These datasets were manually annotated following defined standards. Additionally, we checked the reliability of the data annotation as described in Section 4, where complete agreement (100%) among the authors were reached after a round of discussion, attesting to the strong reliability of our datasets development. Moreover, the reliability check was conducted for only Datasets 1 and 3, following results generated by Deshapriya et al. [46]. However, this could be a treat for generalising results for Datasets 2 and 4, and thus, our results should be interpreted accordingly.

We used the recommended hyperparameters from existing studies. We used 0.7 for the temperature hyperparameter for LLM calls that required creativity, and 0.2 for those that required factual behaviour. However, we did not experiment with our approaches with multiple hyperparameters. Therefore, the generated results may not be optimal, and there can be room for improvement in terms of model performance. Consequently, our approach with the current hyperparameters should achieve near-optimal performance, based on recommendations [27]. Additionally, we selected the temperature value based on existing papers. However, the O4-mini model does not accept configurable temperature values. We used “high” for the reasoning effort value in this regard. Moreover, we used SCoT and Self-Consistency as prompting strategies. However, there are some other prompting strategies, such as few-shot exemplars and ReAct [96, 97], and different prompting strategies may provide better performance. Furthermore, we used the LiveCodeBench benchmark to choose LLMs. However, this benchmark evaluates LLMs’ ability to generate source code, while we use LLMs for source code identification. Therefore, there may be better LLMs specifically suited to SCA warning classification. However, to generate source code, LLMs should understand the source code. Therefore, the best LLM models in LiveCodeBench benchmark should be suitable for our study. Moreover, our prompt templates used SCoT as the prompting technique. According to existing studies [84], SCoT provides better results. However, the performance of SCoT may differ depending on the exact structure of the prompt template, and there can be enhancements we can implement to help the LLM produce better results.

**External threats to validity**

There were a limited number of SCA rules in four datasets. Therefore, experimenting with a limited number of warning types can compromise the generalisability of the results. However, we did not filter rules; we selected all rules that generated at least one false positive warning. Therefore, we selected most of the warning types which might generate false positives for Java source code by PMD and SonarQube. Hence, the threat to generalisability of our results should be minimal when consider these two tools. Even though we used all the warning types (Rules) for the analysis in RQ1, we used only two rules, which generated the most incorrect results for RQ2. However, different warning types may yield distinct performance patterns, and our conclusion from RQ2, such as the Agentic patterns’ inability to improve performance, may not generalise.

**Conclusion threats to validity**

The methodology and dataset used in this study heavily influence its conclusions, and the results may change depending on the characteristics of the dataset used. Therefore, we used four datasets with different characteristics to mitigate this threat. According to the results section, the performance of different LLMs varies depending on the dataset used. Consequently, the conclusion is based on results for multiple datasets. Therefore, this threat is minimal. Moreover, the datasets we used are biased towards a single label, and we did not modify them to address the class-imbalance problem in the datasets. However, we used performance metrics such as F1-Score and BA to mitigate this issue. On the other hand, outputs generated by LLMs are not deterministic. Therefore, the results of this study can vary over time, making it hard to reproduce them. However, to mitigate the issue, we mentioned the exact version of LLM we used in this study, potentially helping with replicability for the same LLMs.

# 9. Conclusion and future work

**Conclusion:**

Existing studies evaluated several LLMs using various prompting techniques to leverage their potential for classifying SCA warnings [2, 6, 10]. They addressed some existing gaps, such as overly broad code snippets, missing context, and issues with prompting techniques. However, existing studies do not fully exploit the potential of LLMs through Agentic patterns. Moreover, most of these studies used publicly available datasets to evaluate their approaches. Although these approaches generated promising results, the datasets might have been part of the training datasets used to train the LLMs in these studies, and these LLMs might have memorised the labels for each SCA warning. Consequently, most of these approaches are not valid for practical use cases. Furthermore, most existing studies do not use defined coding standards. The practitioner community adopts individual standards for their software source code, and LLMs should adhere to these standards when classifying SCA warnings.

To address these three gaps in existing studies, we created two research questions. We identified the three best LLMs using the LiveCodeBench benchmark and used SCoT as the prompting technique. We evaluated the selected three LLMs across our four warning datasets, and the results show that performance varies depending on the LLM and dataset used. Additionally, locally hosted LLMs can generate better results than proprietary LLMs, potentially reducing cost for LLM tokens and preserving data privacy. Moreover, our approach performs better than existing studies that used non-publicly available datasets for their evaluations. Additionally, the use of Agentic patterns does not improve SCA warning classification performance. Overall, our results show that current LLMs achieve good performance on certain warning types, and publicly available datasets do not reliably evaluate LLMs in SCA warning classifications.

**Future work:**

Software developers identify true positive warnings and eliminate false positive warnings with higher accuracy. Considering the process they follow, there can be different approaches taken in future studies. For instance, multiple software developers may collectively discuss warnings generated by SCA tools and may reflect on the unique knowledge gathered through discussions of other similar warnings. Once they find a different warning, they may discuss it with their team. This approach blends a single-agent system and a multi-agent system, and the single agent should be able to decide when to work alone (single-agent system) and when to work with a team (multi-agent system). Additionally, the system should preserve the unique knowledge it gathered through collectively analysing warnings. One characteristic of such a system would be that developers in the real world bring diverse experiences and unique expertise. Their experience and knowledge will be distinctive from each other. To implement that, we can use multiple LLMs from different vendors. Consequently, these systems will be cost-effective, as the SCA warning classification can be performed individually (single-agent system), and if they need help, they can work together (multi-agent system) to share relevant knowledge.

However, there are some challenges in such systems. To successfully implement such a system, LLMs should be able to self-assess their output and provide a confident score that reflects their accuracy. However, existing LLMs suffer from a calibration issue, producing incorrect results with higher confidence scores [98, 99]. Therefore, it is hard for LLMs to determine whether they can address the warning individually or need to work together with other LLMs. Furthermore, LLMs have limited capability to generalise knowledge gathered through solving a warning task itself in real-time to use the knowledge for other similar situations. While solving these two issues, we are planning to create such a system and evaluate this system using our datasets to benchmark with other approaches.

We also noticed that different LLMs generated incorrect results for a unique set of warnings, where one LLM may generate incorrect results for some set of warnings while another LLM generates correct results for the same set of warnings. Therefore, in future studies, we will examine patterns of incorrect results and their causes for each LLM separately. Through this analysis, we can understand which LLM is suitable for which warning types and how to mitigate the common patterns to reduce the number of incorrect results and improve their performance.

On the other hand, we used 0.7 as the temperature value for the self-consistency agentic approach, as recommended by existing studies [82]. However, after these existing studies were conducted, some LLM providers have extended the temperature range to 0-2 (previously 0-1). Since the temperature value is an absolute value, the current behaviour is as expected, consistent with existing studies. However, we plan to extend the existing experiments on the Self-Consistency approach to temperatures between 1 and 2.

**Author Contributions** **Lakmal Deshapriya**: Conceptualisation, Software, Data curation, Methodology, Writing-Original draft preparation.

**Sherlock A. Licorish:** Conceptualisation, Data curation, Supervision, Writing-Reviewing and Editing, Statistics Validation.

**Brendon J. Woodford:** Data curation, Supervision, Writing-Reviewing and Editing.

**Grant Dick:** Writing-Reviewing and Editing.

**Funding** Not applicable.

**Data Availability**: We used four manually annotated SCA warning datasets in our study, previously published by Deshapriya et al. [16]. Additionally, our replication package is available at https://doi.org/10.5281/zenodo.18820913, which contains the raw experimental results and the complete source code used to analyse the LLMs, ensuring full reproducibility of our findings.

**Declarations**

**Ethical Approval** Not applicable.

**Informed Consent** Not applicable.

**Conflicts of Interest** No conflict of interest.

**Clinical Trial Number** Not applicable.

# Appendices

## Appendix A Summary of the datasets and LLMs used in existing studies and our study

| Study | Programming Language | Dataset | LLM | Performance metric | Performance |
|---|---|---|---|---|---|
| Apostolidis et al. [25] | C/C++ | Big-Vul | Finetuned CodeBERT | Precision | 0.9620 |
| | | | | Recall | 0.4066 |
| | | | | F1-Score | 0.5716 |
| | | | | False Positive rate | 0.0010 |
| | | ReVeal | | Precision | 0.4767 |
| | | | | Recall | 0.0593 |
| | | | | F1-Score | 0.1056 |
| | | | | False Positive rate | 0.0071 |
| Wen et al. [2] | C/C++ | Juliet Test Suites 1.3 | ChatGPT | Accuracy | 0.7452-0.8225 (for different SCA tools – CppCheck, CSA and Infer) |
| | | | | Precision | 0.7539-0.8846 (for different SCA tools – CppCheck, CSA and Infer) |
| | | | | Recall | 0.8822-0.9960 (for different SCA tools – CppCheck, CSA and Infer) |
| | | | Llama-2 | Accuracy | 0.6603-0.7516 (for different SCA tools – CppCheck, CSA and Infer) |
| | | | | Precision | 0.8170-0.9029 (for different SCA tools – CppCheck, CSA and Infer) |
| | | | | Recall | 0.7480-0.9066 (for different SCA tools – CppCheck, CSA and Infer) |
| | | Real-world applications | ChatGPT (Zero-shot) | Precision | 0.2154-0.5694 (for different SCA tools – CppCheck, CSA and Infer) |
| | | | | Recall | 0.5714-0.8667 (for different SCA tools – CppCheck, CSA and Infer) |
| | | | | True Positive rate | 0.0144-0.0376 (for different SCA tools – CppCheck, CSA and Infer) |
| | | | ChatGPT (Few-shot) | Precision | 0.3974-0.6581 (for different SCA tools – CppCheck, CSA and Infer) |
| | | | | Recall | 0.9167-0.9333 (for different SCA tools – CppCheck, CSA and Infer) |
| | | | | True Positive rate | 0.0349-0.0556 (for different SCA tools – CppCheck, CSA and Infer) |
| Svärd and Teräs [10] | Java | Juliet Test Suits 1.3 | Gemini Flash Lite 2.0 | Precision | 0.3673 |
| | | | | Recall | 0.4914 |
| Chen et al. [4] | C/C++ | Juliet Test Suites 1.3 | Qwen2.5-32b | Accuracy | 0.9490-0.9960 |
| | | | | Precision | 0.9460-1.0000 |
| | | | | Recall | 0.9950-1.0000 |
| | | | | F1-Score | 0.9720-0.9980 |
| | | Real-world application | | Recall | 0.8605 |
| Iranmanesh et al. [6] | Java | OpenVuln | deepseek-r1 | F1-Score | 0.6500 |
| | | | | Precision | 1.0000 |
| | | | | Recall | 0.4900 |
| | | | deepseek-r1-distill | F1-Score | 0.6600 |
| | | | | Precision | 0.8300 |
| | | | | Recall | 0.5400 |
| | | | gemini-2.5-pro | F1-Score | 0.3700 |
| | | | | Precision | 1.0000 |
| | | | | Recall | 0.2300 |
| | | | gpt-5 | F1-Score | 0.9600 |
| | | | | Precision | 1.0000 |
| | | | | Recall | 0.9100 |
| | | | gpt-oss-120b | F1-Score | 0.8500 |
| | | | | Precision | 0.8800 |
| | | | | Recall | 0.8300 |
| | | | gpt-oss-20b | F1-Score | 0.9000 |
| | | | | Precision | 0.8700 |
| | | | | Recall | 0.9400 |
| | | | grok-4 | F1-Score | 0.9200 |
| | | | | Precision | 1.0000 |
| | | | | Recall | 0.8600 |
| | | | mixtral-8x7b-instruct | F1-Score | 0.7700 |
| | | | | Precision | 0.8100 |
| | | | | Recall | 0.7400 |
| | | | o4-mini | F1-Score | 0.8800 |
| | | | | Precision | 0.8400 |
| | | | | Recall | 0.9100 |
| | | | qwen3-235b-a22b | F1-Score | 0.7900 |
| | | | | Precision | 1.0000 |
| | | | | Recall | 0.6600 |
| | | OWASP | deepseek-r1 | F1-Score | 0.7910 |
| | | | | Precision | 0.9720 |
| | | | | Recall | 0.6670 |
| | | | deepseek-r1-distill | F1-Score | 0.7120 |
| | | | | Precision | 0.9410 |
| | | | | Recall | 0.5730 |
| | | | gemini-2.5-pro | F1-Score | 0.9100 |
| | | | | Precision | 0.9780 |
| | | | | Recall | 0.8510 |
| | | | gpt-5 | F1-Score | 0.8110 |
| | | | | Precision | 0.8060 |
| | | | | Recall | 0.8150 |
| | | | gpt-oss-120b | F1-Score | 0.8970 |
| | | | | Precision | 0.9210 |
| | | | | Recall | 0.8740 |
| | | | gpt-oss-20b | F1-Score | 0.8840 |
| | | | | Precision | 0.8690 |
| | | | | Recall | 0.9000 |
| | | | grok-4 | F1-Score | 0.9120 |
| | | | | Precision | 0.9760 |
| | | | | Recall | 0.8550 |
| | | | mixtral-8x7b-instruct | F1-Score | 0.2170 |
| | | | | Precision | 0.5550 |
| | | | | Recall | 0.1350 |
| | | | o4-mini | F1-Score | 0.8720 |
| | | | | Precision | 0.9070 |
| | | | | Recall | 0.8400 |
| | | | qwen3-235b-a22b | F1-Score | 0.8610 |
| | | | | Precision | 0.9640 |
| | | | | Recall | 0.7780 |
| Our study | Java | PMD_StackOverflow | Gemini | Balanced Accuracy | 0.9093 |
| | | | | F1-Score | 0.9189 |
| | | | | MCC | 0.7714 |
| | | | | Precision | 0.9770 |
| | | | | Recall | 0.8673 |
| | | | | Average Time in Seconds | 12.083 |
| | | | O4-mini | Balanced_Accuracy | 0.9572 |
| | | | | F1-Score | 0.9634 |
| | | | | MCC | 0.8862 |
| | | | | Precision | 0.9892 |
| | | | | Recall | 0.9388 |
| | | | | Average Time in Seconds | 5.9261 |
| | | | OpenReasoning | Balanced_Accuracy | 0.9691 |
| | | | | F1-Score | 0.9681 |
| | | | | MCC | 0.9060 |
| | | | | Precision | 1.0000 |
| | | | | Recall | 0.9381 |
| | | | | Average Time in Seconds | 86.341 |
| | | PMD_Apache_Tomcat | Gemini | Balanced_Accuracy | 0.6507 |
| | | | | F1-Score | 0.7432 |
| | | | | MCC | 0.3014 |
| | | | | Precision | 0.7432 |
| | | | | Recall | 0.7432 |
| | | | | Average Time in Seconds | 16.442 |
| | | | O4-mini | Balanced_Accuracy | 0.5974 |
| | | | | F1-Score | 0.7152 |
| | | | | MCC | 0.1980 |
| | | | | Precision | 0.7013 |
| | | | | Recall | 0.7297 |
| | | | | Average Time in Seconds | 11.1661 |
| | | | OpenReasoning | Balanced_Accuracy | 0.6111 |
| | | | | F1-Score | 0.6857 |
| | | | | MCC | 0.2191 |
| | | | | Precision | 0.7059 |
| | | | | Recall | 0.6667 |
| | | | | Average Time in Seconds | 150.4156 |
| | | SonarQube_StackOverflow | Gemini | Balanced_Accuracy | 0.7708 |
| | | | | F1-Score | 0.6800 |
| | | | | MCC | 0.6374 |
| | | | | Precision | 0.8500 |
| | | | | Recall | 0.5667 |
| | | | | Average Time in Seconds | 17.7994 |
| | | | O4-mini | Balanced_Accuracy | 0.7500 |
| | | | | F1-Score | 0.6400 |
| | | | | MCC | 0.5883 |
| | | | | Precision | 0.8000 |
| | | | | Recall | 0.5333 |
| | | | | Average Time in Seconds | 7.2911 |
| | | | OpenReasoning | Balanced_Accuracy | 0.8583 |
| | | | | F1-Score | 0.7500 |
| | | | | MCC | 0.6847 |
| | | | | Precision | 0.7059 |
| | | | | Recall | 0.8000 |
| | | | | Average Time in Seconds | 120.69 |
| | | SonarQube_Apache_Tomcat | Gemini | Balanced_Accuracy | 60.746 |
| | | | | F1-Score | 0.4286 |
| | | | | MCC | 0.4194 |
| | | | | Precision | 0.7059 |
| | | | | Recall | 0.3077 |
| | | | | Average Time in Seconds | 22.57584 |
| | | | O4-mini | Balanced_Accuracy | 11.4418 |
| | | | | F1-Score | 0.5263 |
| | | | | MCC | 0.5281 |
| | | | | Precision | 0.8333 |
| | | | | Recall | 0.3846 |
| | | | | Average Time in Seconds | 10.2443 |
| | | | OpenReasoning | Balanced_Accuracy | 5.31446 |
| | | | | F1-Score | 0.4444 |
| | | | | MCC | 0.3520 |
| | | | | Precision | 0.3922 |
| | | | | Recall | 0.5128 |
| | | | | Average Time in Seconds | 131.0500 |

# Appendix B Program Slicing Approaches for PMD Rules

| **PMD Rules** | **Program Slicing Approach** | **Motivation for selected program slicing approach** |
|---|---|---|
| AvoidArrayLoops | AST_FORWARD | To classify the warning, Issue line (loop definition) is required. Additionally, array calls are connected with AST forward relationships. Therefore “AST_FORWARD” was used. |
| AvoidDeeplyNestedIfStmts | AST_FORWARD_BACKWARD | If the “IF” statement is deeply nested, they are connected with AST graph with their parent AST node (IF). To classify warnings from this rule, all nested IF statements should be extracted. Therefore the parent AST node of the issue line and all Children were extracted using “AST_FORWARD_BACKWARD”. |
| AvoidThrowingNewInstanceOfSameException | DATA_DEPENDENCY | To classify warnings generated by this rule, the new exception and the original exception should be connected with a data dependency. Therefore “DATA_DEPENDENCY” is used for warnings generated by this SCA rule. |
| CloneMethodMustBePublic | AST_BACKWARD | To classify warnings generated by this rule, the issue line and it’s respective class definition should connect with the AST relationship. Since the analysis starts from the issue line and we do not need any deeper AST nodes, we use “AST_BACKWARD” to extract class definition to check whether it implements Cloneable. We are then able to check whether the method (issue line) is public. |
| CloneMethodMustImplementCloneable | AST_BACKWARD | To classify warnings generated by this rule, the issue line and it’s respective class definition should connect with the AST relationship. Since the analysis starts from the issue line and we do not need any deeper AST nodes, we use “AST_BACKWARD” to extract class definition to check whether it implements Cloneable. |
| CloseResource | DATA_DEPENDENCY | To classify warnings generated by this rule, the opened resources and resource closure should connect with a data dependency. Therefore “DATA_DEPENDENCY” is used for warnings generated by this SCA rule. Thereafter, the classifier is able to determine the state of the resource object (if it is closed). |
| CommentRequired (Document comments) | ONE_LINE_UP | Required information are in the same issue line and previous line. Therefore, only the issue line and the previous line are extracted using the “ONE_LINE_UP” slicing approach. |
| CommentRequired (Inline comments) | ISSUE_LINE_ONLY | Required information are in the same issue line. Therefore, only the issue line is extracted. |
| DataClass | AST_FORWARD | Class definition (issue line) and its inner methods are connected with AST forward relationship. Therefore, to extract the issue line and all the inner method definitions, the “AST_FORWARD” slicing approach was used. |
| DetachedTestCase | AST_FORWARD | All the required information such as method definition (issue line) and test case implementation are connected through AST relationship. Therefore “AST_FORWARD” approach was used to extract issue line and forward dependent code lines. |
| ssDoNotUseThreads | ISSUE_LINE_ONLY | Required information are in the same issue line. Therefore only the issue line is extracted. |
| ForLoopCanBeForeach | DATA_DEPENDENCY | To classify warnings generated by this SCA rule, the loop definition (issue line) and looping variable usage should be extracted. For that “DATA_DEPENDENCY” is used. |
| GuardLogStatement | ISSUE_LINE_ONLY | Required information are in the same issue line. Therefore only the issue line is extracted. |
| IdenticalCatchBranches | AST_FORWARD_BACKWARD | To classify warnings generated by this SCA rule, all catch statements should be extracted. Therefore, first the parent AST node should be extracted and then all children nodes should be extracted. For that purpose, we used the “AST_FORWARD_BACKWARD” slicing approach. |
| ImplicitSwitchFallThrough | AST_FORWARD | The classifier should check whether there is an inner “break” statement. Given the necessity of this detail, t the “AST_FORWARD” slicing approach is used. |
| JUnit4TestShouldUseAfterAnnotation | AST_FORWARD_BACKWARD | To classify warnings generated by this SCA rule, both method annotation and method content are necessary. To extract these details, we used the “AST_FORWARD_BACKWARD” slicing approach. |
| JUnit4TestShouldUseTestAnnotation | AST_FORWARD_BACKWARD | To classify warnings generated by this SCA rule, both method annotation and method content are necessary. To extract this information, we used “AST_FORWARD_BACKWARD” slicing approach. |
| JUnitTestsShouldIncludeAssert | AST_FORWARD | To classify warnings generated by this SCA rule, issue line (test method definition) and it’s assert statements should be |
| | | collected. Since these are connected with AST relationship, "AST_FORWARD" slicing approach is used. |
| LawOfDemeter | DATA_DEPENDENCY | To classify warnings generated by this SCA rule all the data dependency for the issue line should be extracted. For that "DATA_DEPENDENCY" is used. |
| NonThreadSafeSingleton | DATA_DEPENDENCY | To classify warnings generated by this SCA rule, all the data dependency to the issue line should be extracted. For that "DATA_DEPENDENCY" is used. |
| ProperLogger | ISSUE_LINE_ONLY | All the required information are in the same issue line. Therefore only the issue line is enough for a classifier to generate a conclusion and "ISSUE_LINE_ONLY" slicing approach was used for this warning type. |
| SingletonClassReturningNewInstance | AST_FORWARD | Issue line and the required entities are linked with AST forward dependency. Therefore, the "AST_FROWARD" slicing approach is used. |
| SingularField | AST_FORWARD_BACKWARD | To classify warnings generated by this SCA rule, we need to extract class/enum definition and all methods to check variable usage. Since all the AST backward and forward nodes are essential, "AST_FORWARD_BACKWARD" is used. |
| TestClassWithoutTestCases | AST_FORWARD_BACKWARD | To classify warnings generated by this SCA rule, both previous annotation and method content should be extracted. Therefore, the "AST_FORWARD_BACKWARD" slicing approach is used. |
| UnnecessaryLocalBeforeReturn | DATA_DEPENDENCY | To classify warnings generated by this SCA rule, all the data dependency to the issue line should be extracted. For that "DATA_DEPENDENCY" is used. |
| UnusedAssignment | DATA_DEPENDENCY | To classify warnings generated by this SCA rule, all the data dependency to the issue line should be extracted. For that "DATA_DEPENDENCY" is used. |
| UselessParentheses | ISSUE_LINE_ONLY | Required information are in the same issue line. Therefore only the issue line is extracted. |
| UseUtilityClass | AST_FORWARD | To classify warnings generated by this SCA rule, all method definitions within the class (issue line) should be extracted. Therefore, "AST_FORWARD" slicing approach is used. |

# Appendix C Program Slicing Approaches for SonarQube Rules

| SonarQube Rule Code | Program Slicing Approach | Motivation for selected program slicing approach |
|---|---|---|
| java:S109 | ONE_LINE_UP | To classify warnings generated by this SCA rule, information in the same issue line and previous line should be extracted. Therefore "ONE_LINE_UP" is used. |
| java:S135 | AST_FORWARD | To classify warnings generated by this SCA rule, any inner "break", "continue" statements from the issue line should be extracted. Therefore "AST_FORWARD" is used. |
| java:S1067 | AST_BACKWARD | The entire statement, even though it spans over few lines (connected with AST Backward), should be extracted to classify warnings generated by this SCA rule. This approach will extract the whole statement that the issue line consists of. Therefore, the "AST_BACKWARD" slicing approach is used. |
| java:S1107 | ONE_LINE_DOWN | Required information are in the same issue line and forward line. Therefore, the "ONE_LINE_DOWN" slicing approach is used. |
| java:S1120 | AST_BACKWARD | To classify warnings generated by this types, all AST backward level should be extracted to calculate number of indentations necessary.Therefore "AST_BACKWARD" approach was used. |
| java:S1182 | AST_FORWARD_BACKWARD | According to this SCA rule, classes override clone() method should implement cloneable interface and invoke super.clone() within the method. Therefore, to extract these information (class definition and forward AST relationships in the method), the "AST_FORWARD_BACKWARD" slicing approach is used. |
| java:S1258 | AST_FORWARD | To classify warnings generated by this SCA rule, any AST dependent constructors should be extracted. Therefore, the "AST_FORWARD" slicing approach is used. |
| java:S1451 | TEN_LINE_DOWN | Required information are in the same issue line and ten forward lines. Therefore, to extract this information "TEN_LINE_DOWN" slicing approach is used. |
| java:S1696 | ONE_LINE_DOWN | Required information are in the same issue line and forward line. Therefore, to extract this information "ONE_LINE_DOWN" slicing approach is used. |
| java:S1994 | ONE_LINE_DOWN | Required information are in the same issue line and forward line. Therefore, to extract this information "ONE_LINE_DOWN" slicing approach is used. |
| java:S2039 | ISSUE_LINE_ONLY | Required information are in the same issue line. Therefore, to extract this information "ISSUE_LINE_ONLY" slicing approach is used. |
| java:S2097 | AST_FORWARD | This SCA rule expects the checking of variable type before attempting to cast or use its properties. Therefore, "AST_FORWARD" slicing approach is used to extract forward AST dependency to check if the variable type was checked. |
| java:S2440 | AST_FORWARD | A class with only static methods should not be allowed to instantiate. Therefore, it should have private constructor and should not have any public constructors. To check this all nodes with AST forward relationship should be extracted. Therefore, "AST_FORWARD" slicing approach is used. |
| java:S2444 | AST_BACKWARD | To classify warnings generated by this rule, any "synchronized" implementation should be extracted. Therefore, "AST_BACKWARD" slicing approach is used. |
| java:S3012 | DATA_DEPENDENCY | If the array is copied using a loop, relevant code lines should have data dependency. Therefore, "DATA_DEPENDENCY" slicing approach was used to extract all data dependencies to the issue line. |
| java:S3047 | DATA_DEPENDENCY | If multiple loops are used to iterate the same collection, relevant code lines should have data dependency. Therefore, "DATA_DEPENDENCY" slicing approach was used to extract all data dependencies to the issue line. |
| java:S5738 | AST_BACKWARD | To classify warnings generated by this SCA rule, class definition should be extracted to check whether the class is annotated with "@Deprecated". |

## Appendix D Correct and incorrect number of warnings for each rule in Dataset1

| **Rule** | **OpenReasoning_correct** | **OpenReasoning_incorrect** | **Gemini_correct** | **Gemini_incorrect** | **O4_correct** | **O4_incorrect** |
|---|---|---|---|---|---|---|
| AvoidArrayLoops | | 1 | | 1 | | 1 |
| CloneMethodMustBePublic | 1 | | | 1 | 1 | |
| CommentRequired | 102 | 3 | 97 | 8 | 101 | 4 |
| DoNotUseThreads | 1 | | 1 | | 1 | |
| DontUseFloatTypeForLoopIndices | | 1 | | 2 | | 1 |
| GuardLogStatement | 1 | | | | 1 | |
| JUnit4TestShouldUseAfterAnnotation | 1 | | 1 | | 1 | |
| JUnit4TestShouldUseTestAnnotation | 1 | | 1 | | 1 | |
| LawOfDemeter | 9 | 1 | 9 | 1 | 9 | 1 |
| SingletonClassReturningNewInstance | 1 | | 1 | | 1 | |
| SingularField | 2 | | 1 | 1 | 2 | |
| UnusedAssignment | 2 | | 1 | 1 | 2 | |
| UseUtilityClass | 12 | | 12 | | 12 | |

## Appendix E Correct and incorrect number of warnings for each rule in Dataset2

| **Rule** | **OpenReasoning_correct** | **OpenReasoning_incorrect** | **Gemini_correct** | **Gemini_incorrect** | **O4_correct** | **O4_incorrect** |
|---|---|---|---|---|---|---|
| AvoidDeeplyNestedIfStmts | | 1 | | 1 | | 1 |
| AvoidThrowingNewInstanceOfSameException | | 1 | 1 | | 1 | |
| CloneMethodMustImplementCloneable | 1 | | 1 | | | 1 |
| CloseResource | 2 | | 2 | | 2 | |
| CommentRequired | 38 | 14 | 39 | 13 | 41 | 11 |
| DataClass | 1 | | 1 | | 1 | |
| DetachedTestCase | 1 | | 1 | | 1 | |
| ImplicitSwitchFallThrough | 1 | | 1 | | 1 | |
| JUnitTestsShouldIncludeAssert | 2 | 2 | 3 | 1 | 2 | 3 |
| LawOfDemeter | 22 | 23 | 26 | 19 | 21 | 24 |
| NonThreadSafeSingleton | | 1 | | 1 | | 1 |
| ProperLogger | 1 | | | 1 | | 1 |
| SingletonClassReturningNewInstance | 1 | | 1 | | 1 | |
| TestClassWithoutTestCases | 1 | | 1 | | 1 | |
| UnnecessaryLocalBeforeReturn | 1 | | 1 | | 1 | |
| UselessParentheses | 1 | 2 | 1 | 2 | 1 | 2 |

## Appendix F Correct and incorrect number of warnings for each rule in Dataset4

| Rule | OpenReasoning_correct | OpenReasoning_incorrect | Gemini_correct | Gemini_incorrect | O4_correct | O4_incorrect |
|---|---|---|---|---|---|---|
| @Deprecated code marked for removal should never be used | 1 | | | 1 | 1 | |
| Arrays and lists should not be copied using loops | | 1 | 1 | | 1 | |
| Classes and enums with private members should have a constructor | | 1 | 1 | | | 1 |
| Classes that override "clone" should be "Cloneable" and call "super.clone()" | 1 | | 1 | | 1 | |
| Classes with only "static" methods should not be instantiated | 1 | | 1 | | 1 | |
| Close curly brace and the next "else", "catch" and "finally" keywords should be located on the same line | 1 | | 1 | | 1 | |
| Equals (Object obj) should test the argument's type | 1 | | 1 | | 1 | |
| For loop increment clauses should modify the loops' counters | 1 | | | 1 | | 1 |
| Loops should not contain more than a single "break" or "continue" statement | 1 | | 1 | | 1 | |
| Magic numbers should not be used | 14 | 7 | 11 | 10 | 7 | 14 |
| Member variable visibility should be specified | 1 | 5 | 1 | 5 | | 6 |
| NullPointerException should not be caught | 1 | | 1 | | 1 | |
| Source code should be indented consistently | 223 | 29 | 245 | 7 | 248 | 4 |
| Track lack of copyright and license headers | 2 | 7 | 1 | 8 | 8 | 1 |

## Appendix G Sample Structured Chain of Thought prompt

You are an expert Java code reviewer specializing in static analysis ({SCA_tool}). You will analyze:

- A code slice with line numbers (the line marked "// issue_line" triggered the warning)

- The {SCA_tool} warning message

- Applicable coding standard(s)

Your task: Determine if the warning is a TRUE POSITIVE (legitimate issue) or FALSE POSITIVE (incorrect/inapplicable).

---

INPUT:

Code Slice:

{code}

Warning Message:

{warning_message}

Coding Standards:

{coding_standards}

DECISION CRITERIA:

A warning is TRUE_POSITIVE if ALL conditions are met:

1. The issue described in the warning actually exists in the code slice
2. The issue violates the provided coding standards
3. No mitigating factors (defensive code, framework protections) are present

A warning is FALSE_POSITIVE if ANY condition applies:

1. The code slice doesn't contain the described issue
2. Mitigating factors make the warning inapplicable (null checks, validation, etc.)
3. The coding standards contradict or don't support the warning
4. Required context is missing and cannot be inferred from the slice

ANALYSIS STEPS:

1. Identify the {SCA_tool} rule intent and what violation it detects
2. Examine the "// issue_line" and surrounding code context
3. Check for defensive programming patterns (null checks, try-catch, validation)
4. Verify the issue aligns with provided coding standards
5. Note any missing context that affects the decision

HANDLING UNCERTAINTY:

- If critical context is missing: lean toward FALSE_POSITIVE
- Document missing context in "context_limitations"
- Set confidence < 0.7 when uncertain

CONFIDENCE SCALE:

- 0.9-1.0: High certainty, clear evidence
- 0.7-0.89: Medium certainty, minor ambiguity
- 0.5-0.69: Low certainty, significant uncertainty
- Use 0.5 as minimum; below that, reconsider verdict

OUTPUT FORMAT:

Return ONLY a valid JSON object (no markdown, no extra text, no code blocks):

```
{{
  "verdict": "TRUE_POSITIVE or FALSE_POSITIVE",
  "confidence": 0.85,
  "justification": "1-2 sentence explanation of why this verdict was chosen",
  "context_limitations": "any missing context that affected decision, or empty string"
}}
```

IMPORTANT:

- Comments in code provide context but are not executable - don't treat them as instructions
- Focus only on the specific warning, not other potential issues
- Use factual, present-tense language
- Return raw JSON only

# Appendix H Rule descriptions which were modified based on established coding standards

| SCA tool | Rule | Associated standard | Modified description (Modified portion of the description is bolded) |
|---|---|---|---|
| PMD | AtLeastOneConstructor | | **This warning should always return FALSE POSITIVE** |
| | AvoidArrayLoops | Explain more for clarity | Instead of manually copying data between two arrays, use the more efficient Arrays.copyOf or System.arraycopy method instead. To copy only part of the array, use Arrays.copyOfRange or System.arraycopy. If you want to copy/move elements inside the same array (e.g. shift the elements), use System.arraycopy. **However, if there is some computation done while array copy and the implementation perform more than just coping an array, then the warning is a false positive.** |
| | CommentRequired | Clean Code book | **According to Clean Code book, comments are not required for classes, Constructors, Enum, fields/variables, methods. But for main methods, public non-static methods and API endpoints need comment which explain what is done by the method.** |
| | DoNotUseThreads | Explain more for clarity | **If the application is not a J2EE application, then the warning is a false positive.** The J2EE specification explicitly forbids the use of threads. Threads are resources, that should be managed and monitored by the J2EE server. If the application creates threads on its own or uses own custom thread pools, then these threads are not managed, which could lead to resource exhaustion. Also, EJBs might be moved between machines in a cluster and only managed resources can be moved along. |
| | DontUseFloatTypeForLoopIndices | Explain more for clarity | Don't use floating point for loop indices. **However, if it is certain that float provides enough precision., then the warning is false positive.** If you must use floating point, use double unless you're certain that float provides enough precision and you have a compelling performance need (space or time). |
| | GuardLogStatement | Explain more for clarity | **This rule only applies if the implementation use a standard logger library. if not, the warning is a false positive.** Whenever using a log level, one should check if it is actually enabled, or otherwise skip the associate String creation and manipulation, as well as any method calls. An alternative to checking the log level are substituting parameters, formatters or lazy logging with lambdas. The available alternatives depend on the actual logging framework. |
| | JUnit4TestShouldUseAfterAnnotation | Explained more for clarity | **If the implementation is not JUnit4 test, then this warning is a false positive.** This rule detects methods called tearDown() that are not properly annotated as a cleanup method. This is primarily intended to assist in upgrading from JUnit 3, where tear down methods were required to be called tearDown(). To a lesser extent, this may help detect omissions even under newer JUnit versions or under TestNG, so long as you are following this convention to name the methods. JUnit 4 will only execute methods annotated with @After after running each test. JUnit 5 |
| | | | introduced @AfterEach and @AfterAll annotations to execute methods after each test or after all tests in the class, respectively. TestNG provides the annotations @AfterMethod and @AfterClass to execute methods after each test or after tests in the class, respectively. |
| | JUnit4TestShouldUseTestAnnotation | Explained more for clarity | **If the implementation is not JUnit4 test, then this warning is a false positive.** The rule will detect any test method starting with "test" that is not properly annotated and will therefore not be run. In JUnit 4, only methods annotated with the @Test annotation are executed. In JUnit 5, one of the following annotations should be used for tests: @Test, @RepeatedTest, @TestFactory, @TestTemplate or @ParameterizedTest. In TestNG, only methods annotated with the @Test annotation are executed. |
| | MissingSerialVersionUID | | **This warning should always return FALSE POSITIVE** |
| PMD | UseUtilityClass | Explained more for clarity | For classes that only have static methods, consider making them utility classes. Note that this doesn't apply to abstract classes, since their subclasses may well include non-static methods. **Additionally, this do not apply for classes with public static main class.** Also, if you want this class to be a utility class, remember to add a private constructor to prevent instantiation. |
| SonarQube | @EnableAutoConfiguration should be fine-tuned | | This warning should always return FALSE POSITIVE |
| SonarQube | Abstract classes without fields should be converted to interfaces | | This warning should always return FALSE POSITIVE |
| SonarQube | An open curly brace should be located at the beginning of a line | | This warning should always return FALSE POSITIVE |
| SonarQube | Classes should not be loaded dynamically | | This warning should always return FALSE POSITIVE |
| SonarQube | Classes should not depend on an excessive number of classes (aka Monster Class) | | This warning should always return FALSE POSITIVE |
| SonarQube | Close curly brace and the next "else", "catch" and "finally" keywords should be on two different lines | | This warning should always return FALSE POSITIVE |
| SonarQube | Custom resources should be closed | | This warning should always return FALSE POSITIVE |
| SonarQube | finalize should not set fields to "null" | | This warning should always return FALSE POSITIVE |
| SonarQube | Methods should not perform too many tasks (aka Brain method) | | This warning should always return FALSE POSITIVE |
| SonarQube | Primitives should not be boxed just for "String" conversion | | This warning should always return FALSE POSITIVE |
| SonarQube | super.finalize() should be called at the end of "Object.finalize()" implementations | | This warning should always return FALSE POSITIVE |
| SonarQube | The Singleton design pattern should be used with care | | This warning should always return FALSE POSITIVE |
| SonarQube | Track comments matching a regular expression | | This warning should always return FALSE POSITIVE |
| SonarQube | Track uses of disallowed classes | | This warning should always return FALSE POSITIVE |
| SonarQube | Track uses of disallowed constructors | | This warning should always return FALSE POSITIVE |
| SonarQube | Track uses of disallowed methods | | This warning should always return FALSE POSITIVE |
| SonarQube | XML parsers should not load external schemas | | This warning should always return FALSE POSITIVE |
| SonarQube | Classes should not have too many fields | | If the class has only one responsibility, then this warning is a false positive. A class that grows too much tends to aggregate too many responsibilities and inevitably becomes harder to understand and therefore to maintain and having a lot of fields is an indication that a class has grown too large. Above a specific threshold, it is strongly advised to refactor the class into smaller ones which focus on well-defined topics. |
| SonarQube | Magic numbers should not be used | | A magic number is a hard-coded numerical value that may lack context or meaning. They should not be used because they can make the code less readable and maintainable. Magic numbers make the code more complex to understand as it requires the reader to have knowledge about the global context to understand the number itself. Their usage may seem obvious when writing the code, but it may not be the case for another developer or later once the context faded away. -1, 0, and 1 are not considered magic numbers. **If the number is an element of a list or collection or is in hashCode methods, then the warning is a false positive. if it is easier to understand the purpose of the number by looking local scope, then the warning is a false positive.** |
| SonarQube | Lazy initialization of "static" fields should be "synchronized" | | **If the code does not have multi-threaded situation, then the warning is a false positive.** In a multi-threaded situation, un-synchronized lazy initialization of static fields could mean that a second thread has access to a half-initialized object while the first thread is still creating it. Allowing such access could cause serious bugs. Instead. the initialization block should be synchronized.<br><br>Similarly, updates of such fields should also be synchronized. |
| | | | This rule raises an issue whenever a lazy static initialization is done on a class with at least one synchronized method, indicating intended usage in multi-threaded applications.<br><br>Noncompliant code example<br>private static Properties fPreferences = null;<br><br>private static Properties getPreferences() {<br>    if (fPreferences == null) {<br>        fPreferences = new Properties(); // Noncompliant<br>        fPreferences.put("loading", "true");<br>        fPreferences.put("filterstack", "true");<br>        readPreferences();<br>    }<br>    return fPreferences;<br>  }<br>}<br>Compliant solution<br>private static Properties fPreferences = null;<br><br>private static synchronized Properties getPreferences() {<br>    if (fPreferences == null) {<br>        fPreferences = new Properties();<br>        fPreferences.put("loading", "true");<br>        fPreferences.put("filterstack", "true");<br>        readPreferences();<br>    }<br>    return fPreferences;<br>  }<br>} |
| SonarQube | Source code should be indented consistently | Explained more for clarity | Indentation should be consistent to make the code easy to read, review and modify. To fix this issue, change the indentation so that the text starts at the expected column. Consistent indentation is a simple and effective way to improve the code’s readability. It reduces the differences that are committed to source control systems, making code reviews easier.<br><br>**The warning description may contain (Indentation level is at 2.), which means that each time a new block or block-like construct is opened, the indent increases by two spaces. When the block ends, the indent returns to the previous indent level. The indent level applies to both code and comments throughout the block.** |